\documentclass[%
 aip,
 amsmath,amssymb,
 reprint,%
]{revtex4-1}

\usepackage{graphicx}
\usepackage{dcolumn}
\usepackage{bm}
\usepackage{chemfig}
\usepackage{amsmath}
\usepackage{amssymb}

\usepackage[utf8]{inputenc}
\usepackage[T1]{fontenc}
\usepackage{mathptmx}
\usepackage{etoolbox}
\usepackage{float}
\usepackage{MnSymbol}
\usepackage{ulem}

\newcommand{\Sec}[1]{Sec.\,\ref{#1}}
\newcommand{\Eq}[1]{Eq.\,(\ref{#1})}
\newcommand{\Fig}[1]{Fig.\,\ref{#1}}

\newcommand{\RNum}[1]{\uppercase\expandafter{\romannumeral #1\relax}}
\usepackage{tikz}

\makeatletter
\def\@email#1#2{%
 \endgroup
 \patchcmd{\titleblock@produce}
  {\frontmatter@RRAPformat}
  {\frontmatter@RRAPformat{\produce@RRAP{*#1\href{mailto:#2}{#2}}}\frontmatter@RRAPformat}
  {}{}
}%
\makeatother
\begin{document}

\preprint{AIP/123-QED}

\title{Non-equilibrium origin of cavity-induced resonant modifications of chemical reactivities}
\author{Yaling Ke}
\email{yaling.ke@phys.chem.ethz.ch}
\affiliation{ 
Department of Chemistry and Applied Biosciences, ETH Zürich, 8093 Zürich, Switzerland
}%

\begin{abstract}
In this work, we investigate the influence of light-matter coupling on reaction dynamics and equilibrium properties of a single molecule inside an optical cavity. The reactive molecule is modeled using a triple-well potential, allowing two competing reaction pathways that yield distinct products. Dynamical and equilibrium simulations are performed using the numerically exact hierarchical equations of motion approach in real- and imaginary-time formulations, respectively, both implemented with tree tensor network decomposition schemes.  We consider two illustrative cases: one dominated by slow kinetics and another by ultrafast processes. Our results demonstrate that the rates of ground-state reaction pathways can be selectively enhanced when the cavity frequency is tuned into resonance with a vibrational transition directly leading to the formation of the corresponding product, even when that transition is spectroscopically dark. However, tuning cavity frequency to match an absorption-dominant transition shared across both reaction pathways does not necessarily result in pronounced rate enhancements and selectivity. Together with an additional analysis using an asymmetric double-well model, we highlight the greater complexity of underlying factors governing chemical reactivity, which extend beyond considerations of transition dipole strengths and thermal population distributions that shape linear spectroscopy.
Furthermore, we found that in all scenarios, the equilibrium populations remain unchanged when the molecule is moved into the cavity, regardless of the cavity frequency. Thus, our proof-of-concept study confirms at a fully quantum-mechanical level that cavity-induced modifications of chemical reactivities in resonant conditions arise from dynamical and non-equilibrium interactions between the cavity mode and molecular vibrations, rather than from the significant changes in equilibrium properties. 
\end{abstract}

\maketitle

\section{Introduction}
Strong light-matter coupling has emerged as a powerful tool in recent years,\cite{Ebbesen_2016_ACR_p2403} enabling the modification of material properties such as Bose-Einstein condensation,\cite{Kasprzak_Nat_2006_p409} remote energy transfer,\cite{Coles_Nat.Mater._2014_p712} and charge conductivity.\cite{Orgiu_2015_NM_p1123} Among these advancements, one of the most striking observations is the selective catalysis or suppression of ground-state chemical reactivities--achieved without external laser pumping.\cite{Thomas_2016_ACE_p11634} This effect occurs when reactive molecules are placed inside a microfluidic Fabry-P\'erot cavity and the confined electromagnetic modes are tuned to resonate with specific molecular vibrations. Despite ongoing controversies,\cite{imperatore2021reproducibility,wiesehan2021negligible}  the potential applications of these phenomena to revolutionize chemical synthesis have driven further experimental investigations\cite{Vergauwe_2019_ACIE_p15324,Lather_2019_ACIE_p10635,Thomas_2019_S_p615,Hiura_2019__p,Hirai_2020_ACE_p5370,Pang_2020_ACIE_p10436,Sau_2021_ACIE_p5712,Ahn_Sci_2023_p1165,Patrahau_Angew.Chem.Int.Ed._2024_p202401368} and spurred extensive theoretical studies\cite{Galego_2016_NC_p13841,Galego_2017_PRL_p136001,Galego_2019_PRX_p21057,CamposGonzalezAngulo_2019_NC_p4685,CamposGonzalezAngulo_2023_JCP_p230901,Mandal_2020_JPCL_p9215,Mandal_2022_JCP_p,Mandal_2023_CR_p9786,Yang_2021_JPCL_p9531,Li_2021_JPCL_p6974,Li_2021_NC_p1315,Li_2021_AC_p15661,Sun_2022_JPCL_p4441,Schaefer_2022_NC_p7817,Wang_2022_JPCL_p3317,Fischer_2022_JCP_p154305,Fiechter_2023_JPCL_p8261,Sokolovskii_2023_apa_p,Pavosevic_2023_NC_p2766,Ruggenthaler_2023_CR_p11191,Lindoy_2022_JPCL_p6580,Lindoy_2023_NC_p2733,Lindoy_2024_N_p2617,Ying_2023_JCP_p84104,Hu_2023_JPCL_p11208,Ying_2024_CM_p110,Vega__2024_p,Ke_J.Chem.Phys._2024_p224704,Ke_2024_JCP_p54104,Ke_J.Chem.Phys._2025_p64702} aimed at uncovering the underlying mechanisms, which remain mysterious. 

Experimental analyses at the level of Erying transition state theory have reported remarkable changes in thermodynamic parameters, such as activation free energy and entropy, far exceeding the typical Rabi splitting energy $(\hbar\Omega_{\rm R}<k_{\rm B} T)$ that is a hallmark of the strong coupling regime.\cite{Thomas_2019_S_p615,Thomas_2020_PS_p249,Pang_2020_ACIE_p10436} However, theoretical studies\cite{CamposGonzalezAngulo_2020_JCP_p,Li_2020_JCP_p234107}, reveal that classical transition state theory can not successfully explain the resonant rate modifications observed in experiments. This inconsistency has an important implication: extracting thermodynamic parameters from the conventional Arrhenius plots (logarithm of rates versus inverse temperature) may not necessarily indicate an actual alteration of molecular equilibrium properties due to hybridization with the confined radiation field. 

A recent meticulously designed experiment on switchable spin-crossover complexes provides convincing evidence that resonant strong coupling to Fabry-P\'erot cavity modes has a negligible impact on the electronic ground-state energy landscape.\cite{Hoblos_J.Phys.Chem.C_2025_p3107}  These complexes feature two energetically close electronic ground states with different spin multiplicities, allowing reversible switching through thermal cycling, which exhibits an above-room-temperature phase transition. Since only the low-spin state strongly couples to the cavity mode, \cite{Zhang_J.Phys.Chem.Lett._2023_p6840} any slight perturbation to its ground-state energy level would alter the small energy gap between the two states, manifesting as an evident detectable shift in the transition temperature.  The absence of such shifts beyond measurement precision strongly suggests that electronic ground-state energies remain virtually unaffected. This observation could be rationalized by the vanishingly small light-matter coupling strength per molecule,\cite{MartinezMartinez_ACSPhotonics_2018_p167} which scales as $\Omega_{R}/\sqrt{N}$, where $N$ is a macroscopically large number of molecules in a microcavity.  However, we will demonstrate that even in the single-molecule limit with a non-vanishing light-matter coupling strength, hybridization with the vacuum field does not alter molecular equilibrium properties.

In this work, we employ the numerically exact hierarchical equation of motion (HEOM) approach\cite{Tanimura_1989_JPSJ_p101,Yan_2004_CPL_p216,Ishizaki_J.Phys.Soc.Jpn._2005_p3131,Xu_2007_PRE_p31107,Shi_2009_JCP_p84105,Yan_J.Chem.Phys._2014_p54105,Jin_J.Chem.Phys._2008_p234703,Schinabeck_Phys.Rev.B_2018_p235429,Hsieh_J.Chem.Phys._2018_p14103, Shi_J.Chem.Phys._2018_p174102,Tanimura_2020_JCP_p20901} in both imaginary- and real-time formulations to investigate thermal equilibrium properties and reaction dynamics of a single molecule inside an infrared optical cavity. Our simulations, building on previous dynamical studies conducted within a fully quantum-mechanical framework,\cite{Lindoy_2023_NC_p2733,Lindoy_2024_N_p2617,Ying_2023_JCP_p84104,Hu_2023_JPCL_p11208,Ying_2024_CM_p110,Ke_J.Chem.Phys._2024_p224704,Ke_2024_JCP_p54104,Ke_J.Chem.Phys._2025_p64702} incorporate the noise background stemming from the solvent as well as cavity leakage--factors shown to be critical for capturing the correct rate modification profile as observed in experiments. Notably, when a single molecule couples to external degrees of freedom (DoFs)--such as when it is attached to a metal surface--strong coupling can significantly tilt the potential of mean force at equilibrium away from a symmetric double-well profile, thereby altering the reaction rates.\cite{Ke_2022_JCP_p34103} This raises the question of whether the resonant rate modifications observed in optical cavities using symmetric double-well models\cite{Lindoy_2023_NC_p2733,Lindoy_2024_N_p2617,Ying_2023_JCP_p84104,Hu_2023_JPCL_p11208,Ying_2024_CM_p110,Ke_J.Chem.Phys._2024_p224704,Ke_2024_JCP_p54104,Ke_J.Chem.Phys._2025_p64702} may share a similar equilibrium origin. To address this more generally, we consider a molecular system represented by a reaction coordinate governed by a triple-well potential, which extends beyond the symmetric double-well model and supports two competing reaction routes that yield different products.  Our results reveal that the cavity can selectively catalyze one reaction pathway when a cavity frequency is adjusted in resonance with the relevant vibrational transition, yet leaves the chemical equilibrium unchanged under strong coupling conditions. This finding underscores the fundamentally non-equilibrium nature of resonant rate modifications of chemical reactions in optical cavities, as also suggested by a classical simulation in Ref.\onlinecite{Li_2020_JCP_p234107} and experimentally in Ref.\onlinecite{Lather_2022_CS_p195}.

Morever, our results underscore that chemical reaction dynamics are influenced by factors beyond the thermal population of vibrational states and the strength of their associated transition dipoles that predominantly determine linear absorption spectra. Instead, the intricate dynamical interplay among multiple vibrational transitions and reaction pathways, modulated by the cavity field, also plays a critical role in determining reactivity. Accordingly, the achievement of vibrational strong coupling as observed in linear spectroscopy may not constitute a \textit{sine qua non} for realizing pronounced cavity-induced modifications in chemical reaction rates.\cite{imperatore2021reproducibility,wiesehan2021negligible, muller2024measuring,chen2024exploring} 

The remaining sections of this paper are structured as follows: \Sec{sec:theory} introduces the model Hamiltonian and rate theory based on the reactive flux formalism for a triple-well model system in the condensed phase, which can proceed along two concurrent reaction pathways. We also outline the real-time and imaginary-time HEOM methods for computing dynamics and equilibrium properties, respectively, together with an efficient tree tensor network decomposition for time evolution. \Sec{sec:results} presents numerical results and discusses key findings. Finally, we summarize our conclusions in \Sec{sec:conclusion}, with a generalization to an asymmetric double-well model provided in the Appendix.   

\section{\label{sec:theory}Theory}
\subsection{Model}
\begin{figure}[h]
\centering
  \begin{minipage}[c]{0.45\textwidth}
  \raggedright a) System I
    \includegraphics[width=\textwidth]{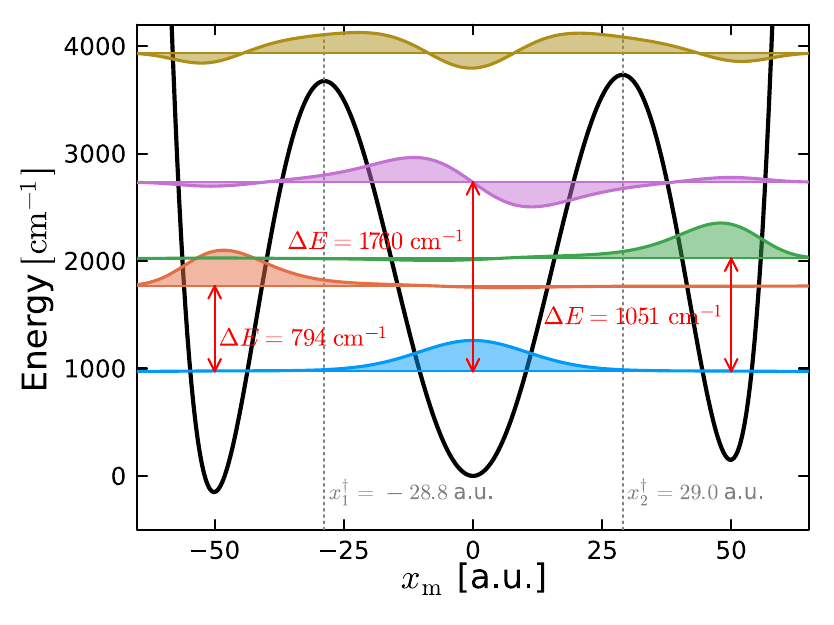}
  \end{minipage}
  \begin{minipage}[c]{0.45\textwidth}
  \raggedright b) System II
    \includegraphics[width=\textwidth]{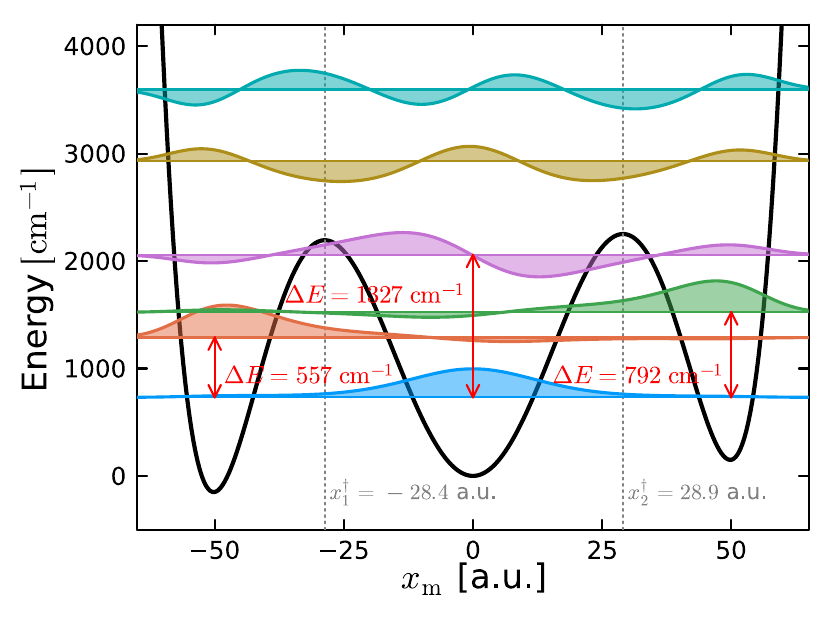}
  \end{minipage}
\caption{Potential energy surface for a triple-well model as defined in \Eq{PES}. The panel a) illustrate the PES with parameters $E_0=25000~{\rm cm}^{-1}$, $a=50~{\rm a.u.}$, and $c=150~{\rm cm}^{-1}$ (System I). The first dividing surface is located $x_{\bf 1}^{\dagger}=-28.8~{\rm a.u.}$ corresponding to a peak with an energy of $3675~{\rm cm}^{-1}$. The second dividing surface is position at $x_{\bf 2}^{\dagger}=29~{\rm a.u.}$, where the peak energy reaches  $3733~{\rm cm}^{-1}$. The panel b) represents the same model but with different parameters: $E_0=15000~{\rm cm}^{-1}$, $a=50~{\rm a.u.}$, and $c=150~{\rm cm}^{-1}$ (System II). Here, the first dividing surface is at $x_{\bf 1}^{\dagger}=-28.4~{\rm a.u.}$ with a peak energy of $2194~{\rm cm}^{-1}$, while the second dividing surface resides at $x_{\bf 2}^{\dagger}=28.9~{\rm a.u.}$ with a peak energy of $2251~{\rm cm}^{-1}$. In both panels, the colored horizontal lines indicate the eigenenergies of the bare molecule, while the wave patterns represent molecular vibrational eigenstates. Additionally, the energy gaps between the vibrational ground state and the first, second, and third excited states are labeled in red.} \label{fig1:PES}
\end{figure}

To study mode-selective chemical reactions in an infrared optical microcavity, we adopt the Pauli-Fierz light-matter Hamiltonian in the dipole gauge under the long-wavelength approximation ($\hbar=1$ is used throughout this work),\cite{Flick_2017_PotNAoS_p3026,Rokaj_2018_JPBAMOP_p34005,Mandal_2023_CR_p9786,Lindoy_2023_NC_p2733} 
\begin{equation}
\label{Hs}
\begin{split}
    H_{\rm S} =  \frac{p_{\rm m}^2}{2} + U(x_{\rm m})  +
    \frac{p_{\rm c}^2}{2} + \frac{1}{2} \omega_{\rm c}^2\left(x_{\rm c} + \sqrt{\frac{2}{\omega_{\rm c}}} \eta_{\rm c} \vec{\mu}(x_{\rm m})\cdot \vec{e} \right)^2.
\end{split}
\end{equation}
Here, $p_{\rm m}$ and $x_{\rm m}$ represent the molecular mass-scaled momentum and reaction coordinate, respectively. The potential energy surface of the electronic ground state, $U(x_{\rm m})$, is modeled as a triple-well function along the reaction coordinate:
\begin{equation}
\label{PES}
    U(x_{\rm m}) = E_{\mathrm{0}}\left(\frac{x_{\rm m}}{a}\right)^2\cdot\left(\left(\frac{x_{\rm m}}{a}\right)^2-1\right)^2+c\left(\frac{x_{\rm m}}{a}\right)^3,
\end{equation}
which defines a potential landscape with three local minima forming distinct wells, as depicted in \Fig{fig1:PES}. The parameters $E_0$ and $a$ control the positions of these minima and the heights of the two energy barriers, while $c$ introduces asymmetry by tilting the potential along the $x_{\rm m}$ axis. The quantized cavity mode is described as a displaced quantum harmonic oscillator with coordinate $q_{\rm c}$, conjugate momentum $p_{\rm c}$, and frequency $\omega_{\rm c}$. The displacement is determined by the light-matter coupling strength, characterized by $\eta_{\rm c}=\frac{1}{\omega_{\rm c}}\sqrt{\frac{\omega_{\rm c}}{2\epsilon_0V}}$, where $\epsilon_0$ is the permittivity of the medium within the cavity, and $V$ denotes the quantization volume of the electromagnetic mode. The unit vector $\vec{e}$ specifies the polarization direction of the cavity field. The molecular dipole moment, $\vec{\mu}(x_{\rm m})$, depends on the reaction coordinate $x_{\rm m}$ and mediates the interaction between the molecule and the confined cavity field.

While the cavity mode in \Eq{Hs} may appear mathematically analogous to a local intramolecular vibrational mode, there are fundamental physical and functional distinctions between the two. Unlike an intrinsic vibrational mode, the cavity mode is an external DoF characterized by its tunable frequency and spatially delocalized nature. These features not only allow precise and flexible experimental control but also enable the potential for inducing collective behavior across multiple molecular systems.

To account for the influence of the solvent as well as ubiquitous far-field electromagnetic modes outside the cavity due to cavity leakage, we consider an open quantum system model, consistent with previous studies.\cite{Fiechter_2023_JPCL_p8261,Lindoy_2023_NC_p2733,Lindoy_2024_N_p2617,Ying_2023_JCP_p84104,Hu_2023_JPCL_p11208,Ying_2024_CM_p110,Ke_J.Chem.Phys._2024_p224704,Ke_2024_JCP_p54104,Ke_J.Chem.Phys._2025_p64702} The total Hamiltonian is given by
\begin{equation}
\label{totalHamiltonian}
    H = H_{\rm S}+ H_{\rm E},
\end{equation}
where $H_{\rm S}$ describes the cavity-molecule system, as defined in \Eq{Hs}, and $H_{\rm E}$ represents its surrounding environment, given by
\begin{equation}
\label{environmentHamiltonian}
\begin{split}
H_{\rm E} = &
 \sum_{k} \frac{P^2_{{\rm m} k}}{2}+\frac{1}{2}\omega_{{\rm m}k}^2 \left(Q_{{\rm m}k}+\frac{g_{{\rm m}k}x_{\rm m}}{\omega_{\rm mk}^2}\right)^2 \\
 &
+\sum_{k'} \frac{P_{{\rm c}k'}^2}{2}+ \frac{1}{2}\omega_{{\rm c}k'}^2 \left(Q_{{\rm c}k'}+\frac{g_{{\rm c}k'}}{\omega_{{\rm c}k'}^2}x_{\rm c}\right)^2.
\end{split}
\end{equation}
Here, we distinguish two types of baths: the solvent bath, which describes numerous solvent DoFs coupled to the molecular reaction coordinate, and the cavity bath, which refers to the continuum of external electromagnetic modes that interact with the confined photonic mode, leading to cavity loss. Each bath comprises an infinite set of harmonic oscillators. Each bath oscillator, indexed by $\alpha$ and $k$, is characterized by its coordinate $Q_{\alpha k}$,  conjugate momentum $P_{\alpha k}$, frequency $\omega_{\alpha k}$, and coupling strength $g_{\alpha k}$, where $\alpha$ denotes either the solvent or cavity bath.    The coupling with the system induces a displacement $g_{\alpha k}x_{\alpha}/\omega_{\alpha k}^2$ in each oscillator's coordinate.  In the absence of such a displacement, the Hamiltonian $H_{\rm E}^0$ represents the environment in isolation. Our recent work has highlighted the crucial role of environmental noise in modulating cavity-induced reaction rate modifications.\cite{Ke_J.Chem.Phys._2025_p64702}

\subsection{Rate theory}
The reactive flux-side correlation function formalism provides a particularly efficient way to evaluate exact quantum-mechanical rate constants for condensed-phase chemical reactions via quantum dynamical simulations.\cite{Miller_1983_JCP_p4889,Craig_2007_JCP_p144503,Chen_2009_JCP_p134505,Ke_2022_JCP_p34103}

In this work, we extend this rate theory--originally developed for a double-well potential--to the more complex case of a triple-well model. The molecular system is divided into three regions: the left, middle, and right wells, separated by two dividing surfaces, $x^{\ddagger}_{\bf 1}$ and $x^{\ddagger}_{\bf 2}$. The first dividing surface $x^{\ddagger}_{\bf 1}$ is positioned at the barrier top between the left and middle wells, while the second, $x^{\ddagger}_{\bf 2}$, resides at the barrier top between the middle and right wells, as illustrated in \Fig{fig1:PES}.
We designate the middle well between two dividing surfaces as the reactant region. The system features two competing reactive pathways, each leading to a distinct product. The first product region is represented by the quantum-mechanical projection operator $1-h_{\bf 1}$,  where $h_{\bf 1}=\theta(x_{\rm m}-x^{\ddagger}_{\bf 1})$ is a Heaviside step function that takes the value 1 when $x_{\rm m}> x^{\ddagger}_{\bf 1}$. The second product region corresponds to the region $x_{\rm m}> x^{\ddagger}_{\bf 2}$ along the reactive coordinate. If both forward and backward reactive processes in two reaction pathways follow a first-order kinetics, they can be described by the scheme:
\begin{equation}
\schemestart
    \chemfig{Product \quad \bf{1}}
    \arrow{<=>[$k_{\rm b, \bf 1}$][$k_{\rm f, \bf 1}$]}
    \chemfig{Reactant}
    \arrow{<=>[$k_{\rm f,\bf 2}$][$k_{\rm b, \bf 2}$]}
    \chemfig{Product \quad \bf 2}.
\schemestop
\end{equation}
The time-dependent populations of two product regions ($P_{\rm p, {\bf 1}/{\bf 2}}(t)$) evolve according to the kinetic equations: 
\begin{subequations}
\label{rateequation}
      \begin{equation}
     \frac{d}{dt} P_{\rm p,\bf 1}(t) = k_{\rm f, \bf 1} P_{\rm r}(t) - k_{\rm b, \bf 1}P_{\rm p, \bf 1}(t),
   \end{equation} 
         \begin{equation}
     \frac{d}{dt} P_{\rm p,\bf 2}(t) = k_{\rm f, \bf 2} P_{\rm r}(t) - k_{\rm b, \bf 2}P_{\rm p, \bf 2}(t), 
   \end{equation} 
\end{subequations}
where $P_{\rm r}(t)$ is the time-dependent population of the reactant. In addition, $k_{\rm f, \bf 1}$ and $k_{\rm b, \bf 1}$ are the thermal rate constants for the forward and backward reactive processes in the first reaction pathway, respectively, while $k_{\rm f, \bf 2}$ and $k_{\rm b, \bf 2}$ represent those for the second pathway. 

At thermodynamic equilibrium, the populations of the reactant and product regions are given by:
\begin{subequations}
    \begin{equation}
        P_{\rm r}^{\rm eq}(\beta) = \rm{tr}\left\{(h_{\bf 1}-h_{\bf 2})\rho^{\rm eq}(\beta) \right\},
    \end{equation}
    \begin{equation}
        P_{\rm p,\bf 1}^{\rm eq}(\beta) = \rm{tr}\left\{(1-h_{\bf 1}) \rho^{\rm eq}(\beta) \right\},
    \end{equation}
    \begin{equation}
        P_{\rm p, \bf 2}^{\rm eq}(\beta) = \rm{tr}\left\{ h_{\bf 2}\rho^{\rm eq}(\beta)\right\},
    \end{equation}
\end{subequations}
where 
\begin{equation}
    \rho^{\rm eq}(\beta) = \frac{1}{Z(\beta)}e^{-\beta H}
\end{equation}
is the canonical equilibrium density operator with $Z(\beta)=\rm{tr}\left\{e^{-\beta H}\right\}$ as the partition function for the overall system as defined in \Eq{totalHamiltonian}, and $\beta=1/k_{\rm B}T$ is the inverse temperature. Since equilibrium populations remain constant, we have $\lim_{t\rightarrow \infty}\frac{d}{dt} P_{\rm p, \bf 1}(t)=\lim_{t\rightarrow \infty}\frac{d}{dt} P_{\rm p, \bf 2}(t)=0$ in the long-time limit where the detailed balance is satisfied:
\begin{equation}
    \frac{k_{\rm f, \bf 1}}{k_{\rm b, \bf 1}} = \frac{P_{\rm p, \bf 1}^{\rm eq}(\beta)}{ P_{\rm r}^{\rm eq}(\beta)} \quad \text{and}   \quad \frac{k_{\rm f, \bf 2}}{k_{\rm b, \bf 2}} = \frac{P_{\rm p, \bf 2}^{\rm eq}(\beta)}{ P_{\rm r}^{\rm eq}(\beta)}.
\end{equation}

Starting from a nonstationary initial state where the system is localized in the reactant region, $\rho_{\rm r}$, the entire system evolves in time according to: 
\begin{equation}
\rho(t) = e^{-iHt} \rho_r e^{iHt}.
\end{equation}
The resulting time-dependent populations of the reactant and two product regions are given by
\begin{subequations}
  \label{population}
\begin{equation}
    P_{\rm r}(t) = \rm{tr}\left\{ (h_{\bf 1}-h_{\bf 2}) \rho(t) \right\},
\end{equation} 
  \begin{equation}
    P_{\rm p, \bf 1}(t) = \rm{tr}\left\{ (1-h_{\bf 1}) \rho(t) \right\},
  \end{equation}
  \begin{equation}
    P_{\rm p, \bf 2}(t) = \rm{tr}\left\{ h_{\bf 2} \rho(t) \right\}.
\end{equation}
\end{subequations}

By combining Eqs.~(\ref{rateequation})-(\ref{population}), we obtain the expressions for the forward rate constants:
\begin{subequations}
\label{rateexpression}
  \begin{equation}
    k_{\rm f, \bf 1} = \lim_{t\rightarrow t_{\rm plateau}} k_{\rm f, \bf 1}(t) = \lim_{t\rightarrow t_{\rm plateau}}\frac{-\rm{tr}\left\{\rho(t)F_{\bf 1} \right\}}{P_{\rm r}(t) -  P_{\rm r}^{\rm eq}(\beta)/P_{\rm p, \bf 1}^{\rm eq}(\beta) \cdot P_{\rm p, \bf 1}(t)},
\end{equation}
\begin{equation}
    k_{\rm f, \bf 2} = \lim_{t\rightarrow t_{\rm plateau}} k_{\rm f, \bf 2}(t) = \lim_{t\rightarrow t_{\rm plateau}}\frac{\rm{tr}\left\{\rho(t)F_{\bf 2} \right\}}{P_{\rm r}(t) -  P_{\rm r}^{\rm eq}(\beta)/P_{\rm p, \bf 2}^{\rm eq}(\beta) \cdot P_{\rm p, \bf 2}(t)}.
\end{equation}
\end{subequations}
Note that \Eq{rateexpression} becomes valid only at long times ($t>t_{\rm plateau} \gg t_{\rm tansient}$), where the transient dynamics subside and the dynamics of the condensed-phase reactive system are governed by rate processes, such that $k_{\rm f,{\bf 1}/{\bf 2}}(t)$ reaches a stationary plateau value.  Here, the flux operator $F_{\bf 1}$ and $F_{\bf 2}$ are the Heisenberg time-derivative of the projection operator $h_{\bf 1}$ and $h_{\bf 2}$, respectively, 
\begin{equation}
    F_{\bf 1} = i[H,h_{\bf 1}] \quad \text{and} \quad F_{\bf 2}=i[H, h_{\bf 2}].
\end{equation}

Thus, the reaction rate constants can be determined by performing exact quantum dynamical and thermodynamic simulations to obtain $\rho(t)$ and $\rho^{\rm eq}(\beta)$, both of which can be efficiently carried out using the HEOM method.

\subsection{HEOM method}
The HEOM method, pioneered by Tanimura and Kubo,\cite{Tanimura_1989_JPSJ_p101} has evolved over the past three decades into a standard non-perturbative and non-Markovian open quantum dynamics approach. It has been widely applied across various fields, ranging from energy transfer in photosynthetic systems to quantum transport in impurity systems.\cite{Yan_2004_CPL_p216,Ishizaki_J.Phys.Soc.Jpn._2005_p3131,Xu_2007_PRE_p31107,Shi_2009_JCP_p84105,Yan_J.Chem.Phys._2014_p54105,Jin_J.Chem.Phys._2008_p234703,Schinabeck_Phys.Rev.B_2018_p235429,Hsieh_J.Chem.Phys._2018_p14103, Shi_J.Chem.Phys._2018_p174102} A comprehensive review of the HEOM method is available in Ref.~\onlinecite{Tanimura_2020_JCP_p20901}.  

\subsubsection{Real-time dynamics}
For an open system coupled to a Gaussian bosonic environment, such as the harmonic baths described in \Eq{environmentHamiltonian}, the system dynamics are described by the reduced system density operator $\rho_{\rm S}(t) = \rm{tr}_{\rm E}\left\{\rho(t)\right\}$, where all harmonic oscillators in the environment are traced out, leaving their influence fully encoded in the bath correlation function
\begin{equation}
\label{timecorrelation}
    C_{\alpha}(t) = \frac{1}{\pi} \int_{-\infty}^{\infty} \frac{e^{-i\omega t}}{1-e^{-\beta \omega}} J_{\alpha}(\omega) \mathrm{d}\omega.
\end{equation}
Here, the spectral density function $J_{\alpha}(\omega)=\frac{\pi}{2}\sum_k \frac{g_{\alpha k}^2}{\omega_{\alpha k}}\delta(\omega -\omega_{\alpha k})$ characterizes the energetic distribution of collective bath modes in a given bath $\alpha$. The reorganization energy, defined as $\lambda_{\alpha}^2=\frac{1}{\pi}\int_0^{\infty}\frac{J(\omega)}{\omega}\mathrm{d}\omega$ quantifies the overall coupling strength between the system and bath $\alpha$. In this work, we assume that the spectral density function for both the solvent and cavity baths takes a Debye–Lorentzian form 
\begin{equation}
\label{Lorentzian}
J_{\alpha}(\omega)=\frac{2\lambda_{\alpha}^2\omega\Omega_{\alpha}}{\omega^2+\Omega_{\alpha}^2},
\end{equation}
where $\Omega_{\alpha}$ is the inverse characteristic correlation timescale of bath $\alpha$. The bath correlation function $C_{\alpha}(t)$ can be expanded analytically or numerically as a sum of exponentials,\cite{Hu_2010_JCP_p101106,Xu_2022_PRL_p230601} 
\begin{equation}
\label{expexpansion}
    C_{\alpha}(t) = \sum_{p=1}^{P\rightarrow \infty} \lambda_{\alpha}^2\eta_{\alpha p} e^{-i\gamma_{\alpha p} t}.
\end{equation}
The explicit expressions of $\eta_{\alpha p}$ and $\gamma_{\alpha p}$ can be found in Ref.~\onlinecite{Ke_2023_JCP_p211102}. This decomposition forms the foundation of the HEOM method, in which a set of auxiliary density operators (ADOs) $\rho^{(n_{{\rm m}1}, \cdots, n_{\alpha p},\cdots, n_{{\rm c} P})}(t)$ is introduced, with each index $n_{\alpha p}$ associated with the $p$-th component in the above exponential expansion. 

Differentiating $\rho^{\mathbf{n}}(t)$ where $\mathbf{n}=(\cdots, n_{\alpha p}, \cdots)$, with respect to time $t$ yields a hierarchy of equations, which take the form 
\begin{equation}
\label{HEOM}
\begin{split}
    i\frac{\mathrm{d}\rho^{\mathbf{n}}(t)}{\mathrm{d}t} = &[H_{\rm S}, \rho^{\mathbf{n}}(t)]-i\sum_{\alpha}\sum_{p} n_{\alpha p} \gamma_{\alpha,p} \rho^{\mathbf{n}}(t) \\
&    + \sum_{\alpha}\sum_p \lambda_{\alpha}\sqrt{ (n_{\alpha p}+1)} \left(x_{\alpha} \rho^{\mathbf{n}_{\alpha p}^+}(t) - \rho^{\mathbf{n}_{\alpha p}^+}(t)x_{\alpha}\right) \\
   & + \sum_{\alpha}\sum_p \lambda_{\alpha}\sqrt{ n_{\alpha p}} \left(\eta_{\alpha p}x_{\alpha} \rho^{\mathbf{n}_{\alpha p}^-}(t) - \eta_{\alpha p}^*\rho^{\mathbf{n}_{\alpha p}^-}(t)x_{\alpha}\right),
\end{split}
\end{equation}
where $\mathbf{n}_{\alpha p}^{\pm}=(\cdots, n_{\alpha p}\pm 1, \cdots)$. In particular, the reduced system density operator is recovered by setting all indices to zero, i.e., $\rho_{S}(t)=\rho^{(0, \cdots, 0,\cdots, 0)}(t)$. 

For an initial factorized state $\rho(t=0)=\rho_{\rm S}(0)\otimes \rho_{\rm E}(0)$, where the environment is in its thermal equilibrium $\rho_{\rm E}(0)=e^{-\beta H_{\rm E}^0}/\rm{tr}_{\rm E}\left\{e^{-\beta H_{\rm E}^0}\right\}$, \Eq{HEOM} can be solved using the explicit initial conditions: $\rho^{\mathbf{0}}(t=0)=\rho_{\rm S}(0)$ and $\rho^{\mathbf{n}}(t=0)=0$ for $||\mathbf{n}||>0$. Since the quantum rate constants (defined in \Eq{rateexpression}) are essentially independent of the initial choice of reactant density operator, we can assume
\begin{equation}
 \rho_r = \frac{1}{Z_r(\beta)}(h_{\bf 1}-h_{\bf 2})e^{-\beta H_{\rm S}} \otimes \rho_{\rm E}(0),   
\end{equation}
where $Z_r(\beta)=\rm{tr}_{\rm S}\left\{ (h_{\bf 1}-h_{\bf 2})e^{-\beta H_{\rm S}} \right\}$ is a partition function traced over the system subspace. Furthermore, the population dynamics in the reactant and two product regions can be extracted from the zeroth-tier ADO as $P_{\rm r}(t) = \rm{tr}_{\rm S}\left\{ (h_{\bf 1}-h_{\bf 2}) \rho^{\mathbf{0}}(t) \right\}$, $P_{\rm p, \bf 1}(t) = \rm{tr}_{\rm S}\left\{ (1-h_{\bf 1}) \rho^{\mathbf{0}}(t) \right\}$, and $P_{\rm p, \bf 2}(t) = \rm{tr}_{\rm S}\left\{ h_{\bf 2} \rho^{\mathbf{0}}(t) \right\}$. The flux correlation functions at two dividing surfaces in \Eq{rateexpression} are calculated through $\rm{tr}\left\{ F_{\bf 1} \rho(t) \right\}=\rm{tr}_{\rm S}\left\{ F_{\bf 1} \rho^{\mathbf{0}}(t) \right\}$ and $\rm{tr}\left\{ F_{\bf 2} \rho(t) \right\}=\rm{tr}_{\rm S}\left\{ F_{\bf 2} \rho^{\mathbf{0}}(t) \right\}$. 

The statistical information of the environment and its correlation with the system are encrypted in the higher-tier ADOs. Therefore, in a more general scenario where the system and environment are entangled at $t=0$, the initial conditions $\rho^{\mathbf{n}}(t=0)$ for propagating \Eq{HEOM} remain implicit, as the higher-tier components $\rho^{\mathbf{n}}(t=0) \neq 0$. In this case, the direct initialization is nontrivial. However, the real-time HEOM method (\Eq{HEOM}) remains solvable when the entire system is in thermal equilibrium. The initial conditions can be determined via the imaginary-time propagation schemes\cite{Tanimura_J.Chem.Phys._2014_p44114,Tanimura_J.Chem.Phys._2015_p144110} or steady-state solvers.\cite{Zhang_J.Chem.Phys._2017_p44105,Kaspar_J.Phys.Chem.A_2021_p5190} 

The system states can be spanned in the eigenstate basis of the bare molecule, denoted as $\{|v_{\rm m}\rangle\}$, which is obtained using the potential-optimized discrete variable representation,\cite{Colbert_1992_JCP_p1982,Echave_1992_CPL_p225} and the harmonic eigenstates of the cavity mode, $\{|v_{\rm c}\rangle$\}. 
To describe the composite system and the effective bath modes (see \Eq{expexpansion}) as a multi-dimensional many-body wavefunction, we employ a twin-space formalism to represent the system in the mixed-state. Furthermore, we introduce second quantization for the effective bath modes using Fock states $|\mathbf{n}\rangle$ in the number representation, along with their associated creation and annihilation operators, $b_{\alpha p}^{+}$ and $b_{\alpha p}$, defined as
\begin{subequations}
\begin{equation}
      b_{\alpha p}^{+} |\mathbf{n}\rangle = \sqrt{n_{\alpha p}+1}| \mathbf{n}_{\alpha p}^+ \rangle;
\end{equation}  
\begin{equation}
      b_{\alpha p} |\mathbf{n}\rangle = \sqrt{n_{\alpha p}}|\mathbf{n}_{\alpha p}^- \rangle,
\end{equation}
\end{subequations}
where $|\mathbf{n}_{\alpha p}^{\pm}\rangle=|\cdots, n_{\alpha p}\pm 1, \cdots\rangle$.  

This formalism enables the construction of an extended wavefunction that encapsulates the entire system--including the molecule, cavity, and the effective bath modes as\cite{Borrelli_2019_JCP_p234102,Borrelli_2021_WCMS_p1539,Ke_2022_JCP_p194102} 
\begin{equation}
\label{Psi}
|\Psi(t)\rangle = \sum_{\bf n}\sum_{v_{\rm m}v'_{\rm m}v_{\rm c}v'_{\rm c}} \mathbb{R}^{\bf n}_{v_{\rm m}v'_{\rm m}v_{\rm c}v'_{\rm c}}| v_{\rm m}v'_{\rm m}v_{\rm c}v'_{\rm c}\rangle \otimes |\mathbf{n}\rangle.
\end{equation}
Here, the coefficient tensor $\mathbb{R}$ connects the density matrix representation of the HEOM to the wavefunction formalism: 
$\mathbb{R}^{\mathbf{n}}_{v_{\rm m}v'_{\rm m}v_{\rm c}v'_{\rm c}}(t)=\langle v'_{\rm c}v'_{\rm m}|\rho^{\mathbf{n}}(t)|v_{\rm m}v_{\rm c}\rangle$. Then, \Eq{HEOM} can be reformulated as a Schr\"odinger-like equation
\begin{equation}
\label{Schroedinger}
    \frac{\mathrm{d}|\Psi(t)\rangle}{\mathrm{d} t} = -i\mathcal{H} |\Psi(t)\rangle,
\end{equation}
where the non-Hermitian super-Hamiltonian $\mathcal{H}$ is explicitly given by 
\begin{equation}
\begin{split}
    \mathcal{H} = &\hat{H}_{\rm S}+\sum_{\alpha}\lambda_{\alpha}^2 \hat{x}_{\alpha}^2 -\tilde{H}_{\rm S}-\sum_{\alpha}\lambda_{\alpha}^2 \tilde{x}_{\alpha}^2 -i \sum_{\alpha}\sum_p  \gamma_{\alpha p} b_{\alpha p}^{+}b_{\alpha p}\\
    & +\sum_{\alpha}\sum_p \lambda_{\alpha}\left[(\hat{x}_{\alpha}-\tilde{x}_{\alpha})b_{\alpha p}     +(\eta_{\alpha p}\hat{x}_{\alpha}-\eta^*_{\alpha p}\tilde{x}_{\alpha})b_{\alpha p}^+ \right].
\end{split}
\end{equation}
Each system operator $O_{j}$ for the $j$th system DoF in Hilbert space is associated with a pair of superoperators in twin space: $\hat{O}_{j}=O_{j}\otimes I_{j}$ and $\tilde{O}_{j}=I_{j}\otimes O_{j}^{\dagger}$, where $I_{j}$ is the identity operator for the $j$th DoF. 

\subsubsection{\label{subsec:imaginary-time}Thermodynamics}
Various thermodynamic variables of a system are determined by the partition function, defined as $Z(\beta) = \rm{tr}\{e^{-\beta H}\}$. However, obtaining $Z(\beta)$ for an open system is highly nontrivial, as the environment typically comprises infinitely many DoFs. Nonetheless, within the imaginary-time HEOM framework, the reduced partition function $Z_{\rm S}(\beta) = Z(\beta)/Z_{\rm E}(\beta)$ can be computed, where $Z_{\rm E}(\beta)= \rm{tr}_E(e^{-\beta H_{\rm E}^0})$, effectively integrating out the environmental DoFs.  The imaginary-time HEOM approach was first established by Tanimura,\cite{Tanimura_J.Chem.Phys._2014_p44114} and several variants have since been developed.\cite{Tanimura_J.Chem.Phys._2015_p144110,Song_J.Chem.Phys._2015_p194106,Ke_J.Chem.Phys._2017_p214105,Zhang_J.Chem.Phys._2022_p174112,Ke_2022_JCP_p34103}

To be more specific, we can introduce a reduced partition operator
\begin{equation}
\label{reducedpartitionoperator}
    \sigma_{\rm S}(\beta) = \frac{1}{Z_{\rm E}(\beta)} \rm{tr}_{\rm E}\left\{e^{-\beta H}\right\}, 
\end{equation}
which yields the reduced partition function after tracing over the system subspace, $Z_{\rm S}(\beta) = \rm{tr}_{\rm S}\left\{  \sigma_{\rm S}(\beta) \right\}$. In the path integral formalism, \Eq{reducedpartitionoperator} is expressed as
\begin{equation}
\begin{split}
\sigma_{\rm S}(\beta) =  & \int \mathcal{D}\bar{x}_{\rm m}(s)\mathcal{D}\bar{x}_{\rm c}(s) e^{-S(\bar{x}_{\rm m}, \bar{x}_{\rm c}, \beta)}  \mathcal{F}(\bar{x}_{\rm m}, \bar{x}_{\rm c}, \beta),
\end{split}
\end{equation}
where $S(\bar{x}_{\rm m}, \bar{x}_{\rm c}, \beta)=\int_0^{\beta} H_{\rm S}(\bar{x}_{\rm m}(s), \bar{x}_{\rm c}(s))\mathrm{d}s$ is the system's Euclidean action functional for the imaginary-time path $\bar{x}_{\alpha}(s)$ evolving from $s=0$ to $s=\beta$. In the presence of the environment, the weight of each path is modified by the influence functional, 
\begin{equation}
\mathcal{F}(\bar{x}_{\rm m}, \bar{x}_{\rm c}, \beta)=e^{\int_0^{\beta}\mathrm{d}s\int_0^{s}\mathrm{d}s' \sum_{\alpha}\bar{x}_{\alpha}(s)G_{\alpha}(s-s')\bar{x}_{\alpha}(s') },    \end{equation}
which depends on the imaginary-time bath correlation function 
\begin{equation}
    G_{\alpha}(s-s')=\frac{1}{\pi}\int_0^{\infty}\mathrm{d}\omega J_{\alpha}(\omega) \frac{\cosh{\left(\beta\omega/2-\omega(s-s')\right)}}{\sinh{\left(\beta\omega/2\right)}}.
\end{equation}
For the Debye-Lorentzian spectral density function given in \Eq{Lorentzian},  $G_{\alpha}(s-s')$ can be expanded as
\begin{equation}
\label{expexpansion_imaginary}
    G_{\alpha}(s-s')=\sum_{p=1}^{P'\rightarrow\infty} \lambda^2_{\alpha}\bar{\eta}_{\alpha p} \cos\left(\bar{\gamma}_{\alpha p}(s-s')\right),
\end{equation}
where $\bar{\gamma}_{\alpha p}=2\pi(p-1)/\beta$, $\bar{\eta}_{\alpha 1}=2/\beta$, and $\bar{\eta}_{\alpha p}=\frac{4\Omega_{\alpha}}{\beta(\Omega_{\alpha}+\bar{\gamma}_{\alpha p})}$ for $p>1$.\cite{Tanimura_J.Chem.Phys._2014_p44114} Building on this expansion, we introduce a series of auxiliary partition operators, 
\begin{equation}
\label{APO}
\begin{split}
\sigma^{\mathbf{m}, \mathbf{m}'}(\tau) =  & \int \mathcal{D}\bar{x}_{\rm m}(s)\mathcal{D}\bar{x}_{\rm c}(s) e^{-S(\bar{x}_{\rm m}, \bar{x}_{\rm c}, \tau)}  \\
&
e^{\int_0^{\tau}ds \int_0^{s}s' \sum_{\alpha, p}\bar{x}_{\alpha}(s) \lambda_{\alpha}^2\bar{\eta}_{\alpha p}\cos \left(\bar{\gamma}_{\alpha p}(s-s')\right)\bar{x}_{\alpha}(s')}  \\
&\prod_{\alpha, p}\frac{1}{\sqrt{m_{\alpha p}!}}\left(\int_0^{\tau}\lambda_{\alpha}\bar{\eta}_{\alpha p}\cos(\bar{\gamma}_{\alpha p}(\tau-s))\bar{x}_{\alpha}(s)\mathrm{d}s\right)^{m_{\alpha p}} \\
&\prod_{\alpha, p}\frac{1}{\sqrt{m'_{\alpha p}!}}\left(\int_0^{\tau}\lambda_{\alpha}\bar{\eta}_{\alpha p}\sin(\bar{\gamma}_{\alpha p}(\tau-s))\bar{x}_{\alpha}(s)\mathrm{d}s\right)^{m'_{\alpha p}}. 
\end{split}
\end{equation}
According to the above definition, at the initial imaginary time $\tau=0$, we have $\sigma^{\mathbf{m=0}, \mathbf{m'=0}}(\beta)=I$, where $I$ is the identity operator in the system subspace, and $\sigma^{\mathbf{m}, \mathbf{m'}}(\tau=0)=0$ for $||\mathbf{m}||+||\mathbf{m}'||>0$.  In addition, at $\tau=\beta$, for $\mathbf{m}=\mathbf{m}'=\mathbf{0}$, we recover the reduced partition operator $\sigma^{\mathbf{m=0}, \mathbf{m'=0}}(\beta)=\sigma_{\rm S}(\beta)$.  Differentiating $\sigma^{\mathbf{m}, \mathbf{m}'}(\tau)$ with respect to $\tau$, we derive the imaginary-time HEOM, given by
\begin{equation}
\label{imHEOM}
\begin{split}
\frac{\partial \sigma^{(\mathbf{m}, \mathbf{m}')}(\tau) }{\partial \tau} =&  -H_s \sigma^{(\mathbf{m}, \mathbf{m}')}(\tau)
+\sum_{\alpha, p}\lambda_{\alpha}\sqrt{m_{\alpha}+1}x_{\alpha} \sigma^{(\mathbf{m}_{\alpha p}^+, \mathbf{m}')}(\tau) \\
&
+\sum_{\alpha, p}\lambda_{\alpha}\sqrt{m_{\alpha p}}\bar{\eta}_{\alpha p}x_{\alpha}\rho^{(\mathbf{m}_{\alpha p}^-, \mathbf{m}')}(\tau) \\
&
+\sum_{\alpha p} \sqrt{m_{\alpha p}+1}\sqrt{m'_{\alpha p}}\bar{\gamma}_{\alpha p}\sigma^{(\mathbf{m}_{\alpha p}^+, \mathbf{m'}_{\alpha p}^-)}(\tau) \\
&
-\sum_{\alpha, p}\sqrt{m_{\alpha p}}\sqrt{m'_{\alpha p}+1}\bar{\gamma}_{\alpha p}\sigma^{(\mathbf{m}_{\alpha p}^-, {\mathbf{m'}^{+}_{\alpha p}})}(\tau). 
\end{split}
\end{equation}

By propagating \Eq{imHEOM} from $\tau=0$ to $\beta$, we can also obtain the reduced system density operator in thermal equilibrium as
\begin{equation}
\rho^{\rm eq}_{\rm S}(\beta)=\frac{\mathrm{tr}_{\rm E}\left\{e^{-\beta H }\right\}}{Z(\beta)} = \frac{\sigma^{(\mathbf{m=0}, \mathbf{m'=0})}(\beta)}{Z_{\rm S}(\beta)},
    \end{equation}
where the reduced partition function is given by $Z_{\rm S}(\beta)=\mathrm{tr}_{\rm S}\{\sigma^{(\mathbf{m=0},\mathbf{m'=0})}(\beta))$. This formulation enables the computation of various thermodynamic properties. For example, we can calculate the equilibrium populations of the reactant and two product regions are given by $P_{\rm r}^{\rm eq}(\beta)=\mathrm{tr}_{\rm S}\left\{(h_{\bf 1}-h_{\bf 2})\rho^{\rm eq}_{\rm S}(\beta)\right\}$, $P_{\rm p,\bf 1}^{\rm eq}(\beta)=\mathrm{tr}_{\rm S}\left\{(1-h_{\bf 1})\rho^{\rm eq}_{\rm S}(\beta)\right\}$, and $P_{\rm p,\bf 2}^{\rm eq}(\beta)=\mathrm{tr}_{\rm S}\left\{h_{\bf 2}\rho^{\rm eq}_{\rm S}(\beta)\right\}$.

Similar to the real-time HEOM method, we can construct an extended partition wavefunction 
\begin{equation}
\label{Psi}
|\Phi(\tau)\rangle \equiv \sum_{\mathbf{m}, \mathbf{m'}} \sum_{v^{}_{\rm m} v'_{\rm m}v_{\rm c} v'_{\rm c}  } \mathbb{W}^{\mathbf{m}, \mathbf{m'}}_{v^{}_{\rm m}v'_{\rm m}v_{\rm c}v'_{\rm c}}(\tau)|v^{}_{\rm m}v'_{\rm m} v_{\rm c}v'_{\rm c}\rangle\otimes|\mathbf{m}\rangle \otimes|\mathbf{m'}\rangle,
\end{equation}
where $\mathbb{W}^{\mathbf{m}, \mathbf{m'}}_{v^{}_{\rm m}v'_{\rm m}v_{\rm c}v'_{\rm c}}(\tau)=\langle v'_{\rm c}v'_{\rm m}|\sigma^{\mathbf{m}, \mathbf{m'}}(\tau)|v_{\rm m}v_{\rm c}\rangle$. The evolution of $|\Phi(\tau)\rangle$ in imaginary time is governed by 
\begin{equation}
    \frac{\mathrm{d}|\Phi(\tau)\rangle}{\mathrm{d}\tau} = - \mathcal{H}_{im}|\Phi(\tau)\rangle
\end{equation}
with the imaginary-time super-Hamiltonian 
\begin{equation}
    \mathcal{H}_{im} = \hat{H}_s  
-\sum_{\alpha, p} \lambda_{\alpha}\hat{x}_{\alpha} \left(b_{\alpha p}
+\bar{\eta}_{\alpha p}b_{\alpha p}^{+}\right) +\bar{\gamma}_{\alpha p}b_{\alpha p}^{+}d_{\alpha p} -\bar{\gamma}_{\alpha p}b_{\alpha p}d^{+}_{\alpha p}.
\end{equation}
Here, the ladder operators $b_{\alpha p}^{\pm}$ and $d_{\alpha p}^{\pm}$ act on the Fock states $|\mathbf{m}\rangle$ and $|\mathbf{m'}\rangle$, respectively, as
\begin{subequations}
\begin{equation}
      b_{\alpha p}^{+} |\mathbf{m}\rangle = \sqrt{m_{\alpha p}+1}| \mathbf{m}_{\alpha p}^+ \rangle \quad \text{and} \quad  d_{\alpha p}^{+} |\mathbf{m'}\rangle = \sqrt{m'_{\alpha p}+1}| \mathbf{m'}_{\alpha p}^+ \rangle, 
\end{equation}  
\begin{equation}
      b_{\alpha p} |\mathbf{m}\rangle = \sqrt{m_{\alpha p}}|\mathbf{m}_{\alpha p}^- \rangle \quad \text{and} \quad     d_{\alpha p} |\mathbf{m'}\rangle = \sqrt{m'_{\alpha p}}|\mathbf{m'}_{\alpha p}^- \rangle.
\end{equation}
\end{subequations}
We note that, by introducing the auxiliary partition operators as defined in \Eq{APO}, the resulting super-Hamiltonian  $\mathcal{H}_{im}$ becomes independent of $\tau$. This independence enables the tensor network decomposition--described below--to be carried out just once prior to time propagation, significantly enhancing the efficiency of equilibirium simulations.

\subsubsection{Tree tensor network states}
To facilitate efficient dynamical and thermodynamical simulations of chemical reactions in optical cavities using the HEOM method, we decompose the high-rank coefficient tensors $\mathbb{R}(t)$ in the real-time extended wavefunction $|\Psi(t)\rangle$ and $\mathbb{W}(\tau)$ in the imaginary-time extended partition wavefunction $|\Phi(\tau)\rangle$ as tree tensor network states (TTNS).

\begin{figure}[h]
\centering
  \begin{minipage}[c]{0.45\textwidth}
  \raggedright a) TTNS
    \includegraphics[width=\textwidth]{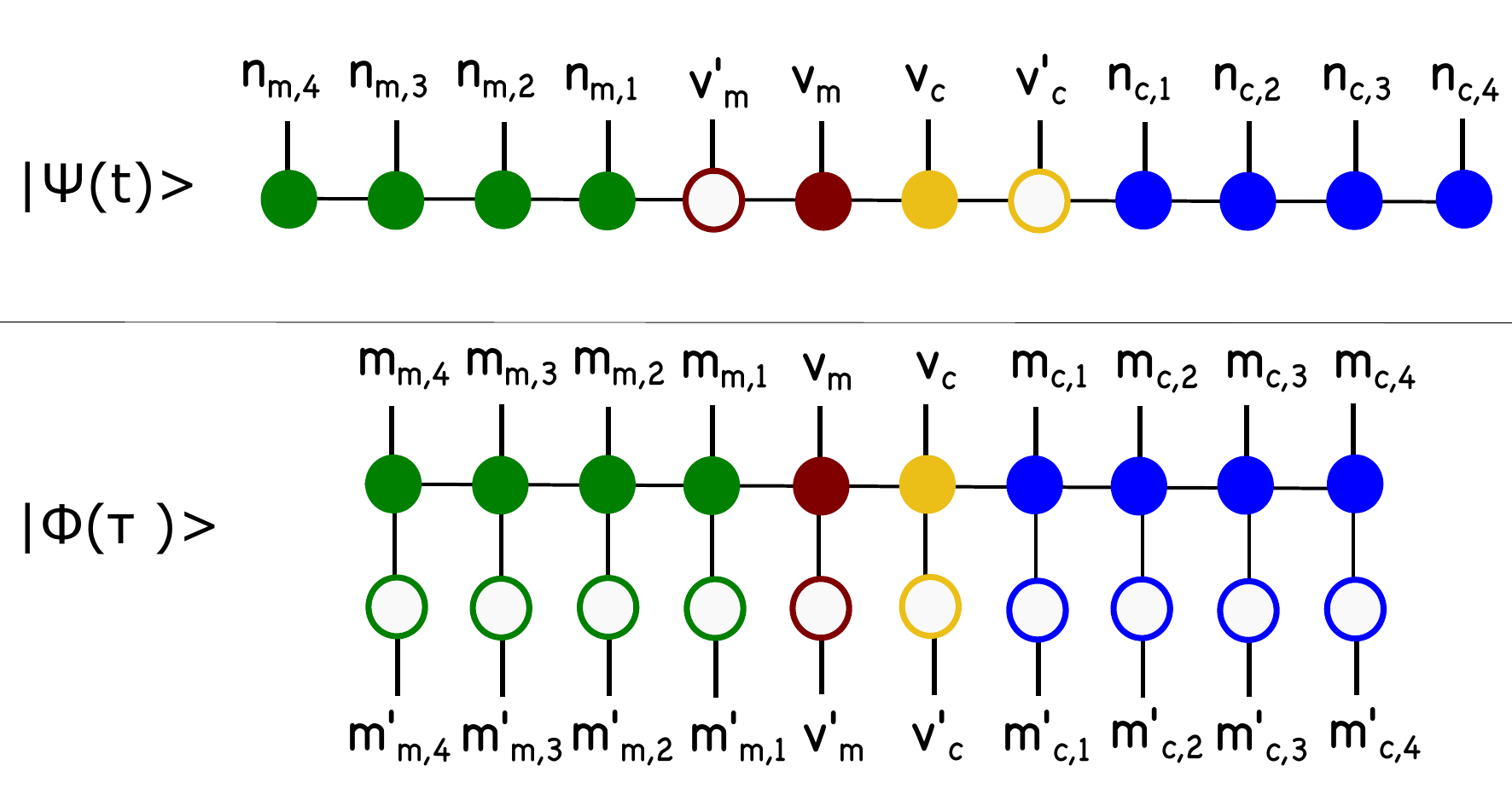}
  \end{minipage}
  \begin{minipage}[c]{0.45\textwidth}
  \raggedright b) TTNO
    \includegraphics[width=\textwidth]{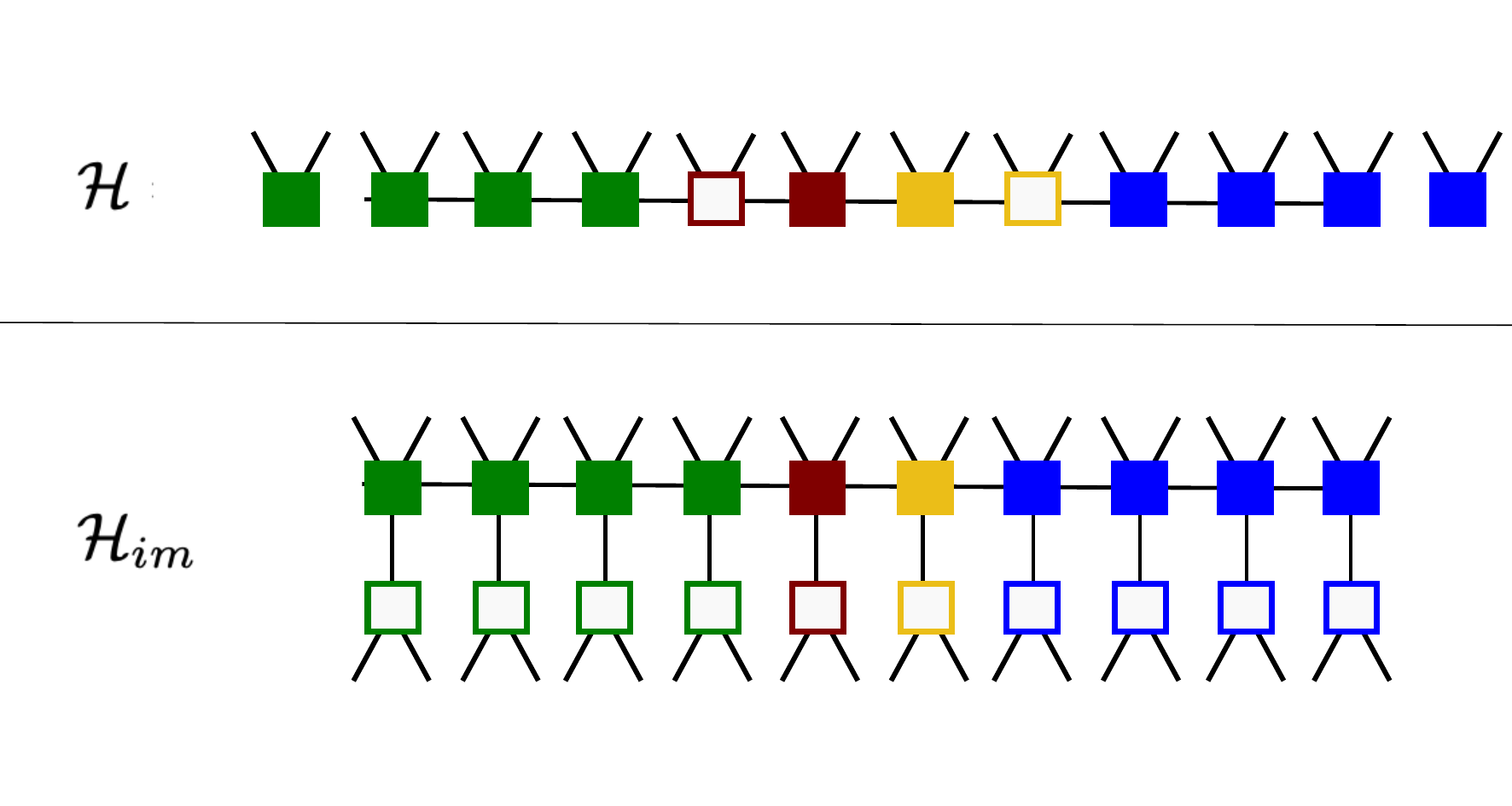}
  \end{minipage}
\caption{ a) Graphical representations of the TTNS decomposition: the top diagram is depicted for the extended wavefunction $|\Psi(t)\rangle$, while the bottom diagram represents the decomposition for the extended partition wavefunction $|\Phi(\tau)\rangle$.  In this example, we set $P=4$ in \Eq{expexpansion} and $P'=4$ in \Eq{expexpansion_imaginary}. Each colored circular node in the TTNS represents a low-rank tensor associated with a specific component of the model: molecule (in red), solvent
(in green), cavity mode (in yellow), and cavity bath (in blue). b) Graphical representations of the TTNO decomposition of the super-Hamiltonian $\mathcal{H}$ at the top and $\mathcal{H}_{\rm im}$ at the bottom. The square nodes represent the local low-rank tensors in the TTNO decomposition, and they share the same color-coding scheme as in
the TTNS. } \label{fig2:TTNS_TTNO}
\end{figure}

Tree tensor network states offer a compact representation for high-dimensional quantum many-body systems, particularly when different DoFs exhibit complex and strong entanglement. In this framework, $\mathbb{R}(t)$ and $\mathbb{W}(\tau)$ are decomposed into a network of interconnected low-rank tensors arranged in a loopless tree topology. After evaluating the computational efficiency of several different tensor network topologies, we choose the matrix product state (MPS), a special instance of TTNS in a one-dimensional chain, for the real-time wavefunction, while adopting a genuine tree structure for the partition wavefunction, as illustrated in \Fig{fig2:TTNS_TTNO}a). This diagram shows nodes with one open leg and two or three connected legs. Each open leg corresponds to a physical dimension. In other words, each node is assigned to a specific DoF. The nodes in red correspond to the reactive molecular DoF, represented by $v_{\rm m}$ (solid) and $v'_{\rm m}$ (hollow). The yellow nodes denote the cavity mode ($v_{\rm c}$ and $v'_{\rm c}$). The green and blue nodes stand for the solvent and cavity bath modes, respectively. A connected leg represents a shared virtual index between two nodes on both sides, which is subject to contraction, and the index runs from $1$ to $D_{i}$. The maximal bond dimension, denoted as $D_{\rm max}$, is the largest value among the bond dimensions $\{D_{i}\}$, and this value is incrementally increased until convergence to a desired level of accuracy is achieved.

Moreover,  we observe that both the super-Hamiltonian $\mathcal{H}$ governing real-time dynamics and the imaginary-time super-Hamiltonian $\mathcal{H}_{\rm im}$ for thermodynamics are expressed as a sum of products. This structure allows an automatic decomposition of the Hamiltonian into an optimal tree tensor network operator (TTNO). This decomposition can be achieved using techniques such as bipartite graph theory optimization\cite{Li_J.Chem.Phys._2024_p54116, cakir2025} or state diagram compression.\cite{Milbradt_SciPostPhys.Core_2024_p36,Milbradt_arXivpreprintarXiv2407.13249_2024_p} In the tensor diagram for TTNOs, as illustrated in \Fig{fig2:TTNS_TTNO}b), each node is represented as a square with two open legs.  By ensuring that the TTNO for the Hamiltonian and the TTNS for the wavefunction share the same tree structure, we can efficiently implement time evolution based on the time-dependent variational principle (TDVP) for time evolution, \cite{Haegeman_2016_PRB_p165116,Paeckel_2019_APN_p167998,Dunnett_2021_PRB_p214302,Bauernfeind_SciPostPhysics_2020_p24,Kloss_SciPostPhysics_2020_p70,Ceruti_SIAMJ.Numer.Anal._2021_p289}. Further technical details are provided in Ref. \onlinecite{Ke_2023_JCP_p211102}.

\section{\label{sec:results}Results}
Using the theoretical methods introduced in \Sec{sec:theory}, we first examine the equibrium properties of symmetric double-well models previously studies in Refs.\,\onlinecite{Fiechter_2023_JPCL_p8261,Lindoy_2023_NC_p2733,Lindoy_2024_N_p2617,Ying_2023_JCP_p84104,Hu_2023_JPCL_p11208,Ying_2024_CM_p110,Ke_J.Chem.Phys._2024_p224704,Ke_2024_JCP_p54104,Ke_J.Chem.Phys._2025_p64702}, confirming that the equlibirium populations within the cavity remain identical to those of the bare molecule. Specifically, the molecule remains equally populated in the reactant and product wells, in contrast to cases where it is coupled to a metal surface, which can significantly alter the potential of mean force and shift equilibrium populations.\cite{Ke_2022_JCP_p34103} To assess the generality of this finding, in the following, we investigate the reaction dynamics and thermal equilibrium properties of a single reactive molecule (System I) described by a triple-well model confined in an optical cavity. The molecule can undergo two distinct reactive pathways, with dynamics governed by slow rate kinetics. We demonstrate mode-selective resonant modifications of reaction rates inside the cavity while observing that the equilibrium populations remain unchanged, irrespective of whether the cavity is tuned on- or off-resonance.  To further illustrate the contrast between resonant dynamical variation in rates and thermodynamic invariance, we examine the same model system but with lower reaction barriers (System II), where the molecular reactive processes occur on an ultrafast timescale.

For our simulations, all calculations are performed using the following parameters: $\lambda_{\rm m}=\lambda_{\rm c}=100~{\rm cm}^{-1}$, $\Omega_{\rm m}=200~{\rm cm}^{-1}$, $\Omega_{\rm c}=1000~{\rm cm}^{-1}$, the ambient temperature $T=300~{\rm K}$. For simplicity, we assume a molecular dipole function of $\mu(x_{\rm m})=x_{\rm m}$, which is perfectly aligned with the cavity field polarization, such that $\mu(x_{\rm m}) \cdot \vec{e}= x_{\rm m}$. The molecular vibrational mode and the cavity mode are represented by $d_{\rm m}$ and $d_{\rm c}$ lowest eigenstates, respectively. With convergence verified for the chosen parameters, unless stated otherwise, we set $d_{\rm m}=10$, $d_{\rm c}=6$, the time step $\delta t = 1~{\rm fs}$, and $\delta \tau = \beta/500$, and the maximal bond dimension $D_{\rm max} = 40$.  Each effective bath mode is represented by $d_{\rm e}=10$ states, defining the hierarchical depth. The real-time bath correlation function in \Eq{timecorrelation} is expanded using the Pad\'e decomposition scheme with $P=4$. In addition, $P'=4$ is used in \Eq{expexpansion_imaginary}.

\subsection{\label{subsec:modelI}System I: slow rate process}
\begin{figure}
\centering
  \begin{minipage}[c]{0.45\textwidth}
  \raggedright a)
    \includegraphics[width=\textwidth]{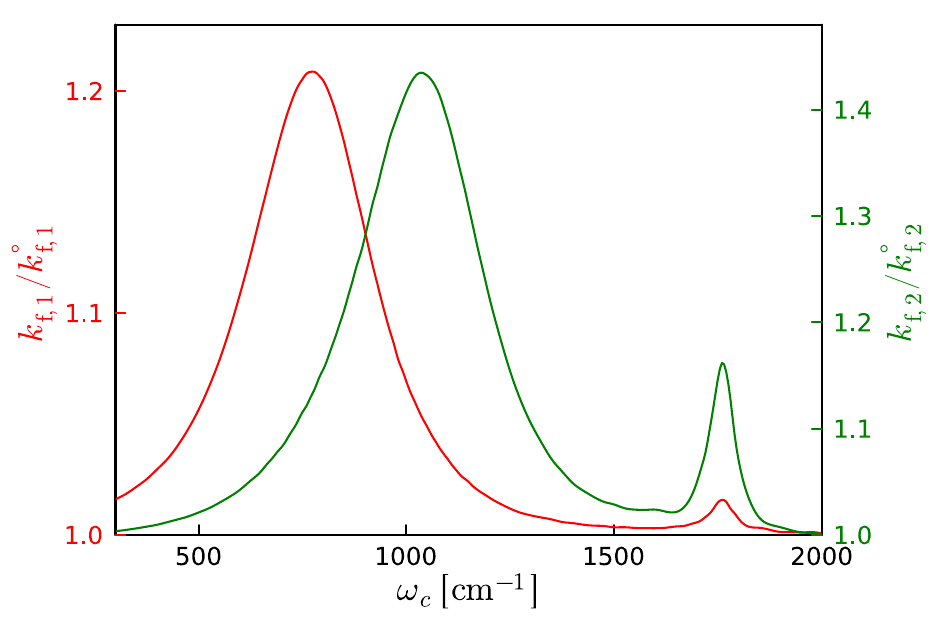}
 \end{minipage}
    \begin{minipage}[c]{0.45\textwidth}
  \raggedright  b) 
    \includegraphics[width=\textwidth]{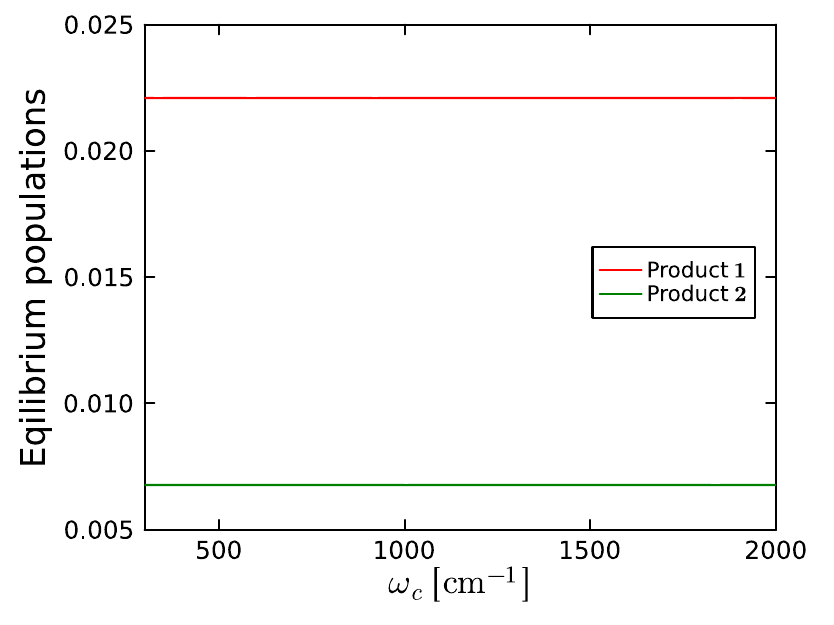}
  \end{minipage}
\caption{a) Ratios of reaction rates inside and outside the cavity for System I as a function of the cavity frequency $\omega_{\mathrm{c}}$. The red and green lines correspond to the first ($k_{\rm f, \bf 1}/k^{\circ}_{\rm f, \bf 1}$, plotted against the left axis) and second ($k_{\rm f, \bf 2}/k^{\circ}_{\rm f, \bf 2}$, plotted against the right axis) reaction pathways, respectively. b) Equilibrium populations of products ${\bf 1}$ and ${\bf 2}$ as a function of $\omega_{\mathrm{c}}$. The light-matter coupling strength is set to $\eta_{\mathrm{c}}=0.00125~$a.u.} \label{fig3:model1}
\end{figure}
We begin with a triple-well model system, where the PES defined in  \Eq{PES} is characterized by the parameters: $E_0=25000~{\rm cm}^{-1}$, $a=50~{\rm a.u.}$, and $c=150~{\rm cm}^{-1}$. The PES and the lowest five eigenstates of the bare molecule are visualized in \Fig{fig1:PES}a). In this system, the reaction barriers from the bottom of the reactant region $x_{\rm m}=0$ to the barrier tops are over an order of magnitude larger than the thermal energy $(\sim 208\mathrm{cm}^{-1})$. Therefore, the chemical reactions are in the deep-tunneling regime, which is not thermally accessible and is dominated by slow rate processes.  

The lowest four vibrational eigenstates lie below the reaction barriers. The vibrational ground and the third excited states are localized primarily in the center reactant region (middle well). The first excited state $|v_{\rm m}=1\rangle$ resides predominantly in the left well (the first product region) and exhibits a transition energy of $\Delta E_{0\rightarrow1}=794~{\rm cm}^{-1}$ from the ground state $|v_{\rm m}=0\rangle$. Hence, excitation to the first excited state initiates the first reaction pathway, producing product ${\bf 1}$.  In contrast, the transition to the second vibrationally excited state $|v_{\rm m}=2\rangle$ leads to a competing reaction pathway, mainly populating in the right well (product ${\bf 2}$), with a transition energy of $\Delta E_{0\rightarrow 2}=1051~{\rm cm}^{-1}$, well separated from that of the first reaction pathway. Notably, the molecular absorption spectrum features a dominant peak centered at approximately $1770\,\mathrm{cm}^{-1}$ (data not shown), corresponding to the vibrational transition $|v_{\rm m}=0\rangle \rightarrow |v_{\rm m}=3\rangle$. This transition is associated with a larger transition dipole matrix element $\langle v_{\rm m}=3|x| v_{\rm m}=0\rangle=7.71$, as compared to those of the lower-energy transitions, such as $\langle v_{\rm m}=1|x| v_{\rm m}=0\rangle=1.35$ and $\langle v_{\rm m}=2|x| v_{\rm m}=0\rangle=1.36$, thereby accounting for its spectral dominance.

When the reactant is placed outside the cavity, oscillatory transient dynamics dominate the short-time regime, lasting up to a few picoseconds. After this, the reaction proceeds through rate processes, indicated by the emergence of a plateau region in $k_{\rm f,\bf 1}(t)$ and $k_{\rm f, \bf 2}(t)$, as defined in \Eq{rateexpression}. 
By incrementally expanding the vibrational basis size $d_{\rm m}$ and tracking the corresponding forward reaction rates  $k_{\rm f, \bf 1}(d_{\rm m})$ and $k_{\rm f, \bf 2}(d_{\rm m})$, we gain deeper microscopic insights into the underlying reactive mechanisms. For the first reaction pathway, two distinct channels are identified: a dominant one-step transition $|v_{\rm m}=0\rangle\rightarrow |v_{\rm m}=1\rangle$ and a supplementary two-step process $|v_{\rm m}=0\rangle\rightarrow |v_{\rm m}=3\rangle\rightarrow |v_{\rm m}=1\rangle$. The one-step transition is the primary contributor, as evidenced by the nearly converged rate $k_{\rm f, \bf 1}(d_{\rm m}=3)=4.9\times10^{-7}~{\rm fs}^{-1}$, which excludes the participation of the $|v_{\rm m}=3\rangle$ state and has closely approached the fully converged rate $k^{\circ}_{\rm f, \bf 1}=5.8\times10^{-7}~{\rm fs}^{-1}$. The two-step channel makes a minor contribution to the overall reaction rate, as it involves initial excitation to the third excited state followed by a rate-limiting relaxation to $|v_{\rm m}=1\rangle$. A similar pattern is observed for the second reaction pathway, which proceeds primarily through the direct transition $|v_{\rm m}=0\rangle\rightarrow |v_{\rm m}=2\rangle$, supplemented by a slower route $|v_{\rm m}=0\rangle\rightarrow |v_{\rm m}=3\rangle\rightarrow |v_{\rm m}=2\rangle$. The total converged forward reaction rate for this pathway is $k^{\circ}_{\rm f, \bf 2}=1.34\times10^{-7}~{\rm fs}^{-1}$.  The branching ratio of the two reaction pathways is $\phi^{\circ}=k^{\circ}_{\rm f, \bf 1}/k^{\circ}_{\rm f, \bf 2}=4.3$, meaning the overall reaction predominantly follows the first reaction pathway, favoring the formation of product ${\bf 1}$. While the reaction is far from completion in the real-time simulations propagated to tens of picoseconds using the HEOM method (\Eq{Schroedinger}), the equilibrium populations can be obtained directly from the imaginary-time evolution, as elaborated in \Sec{subsec:imaginary-time}. The equilibrium population probabilities for products ${\bf 1}$ and ${\bf 2}$ regions are 0.022 and 0.0068, respectively, yielding a ratio of 3.3, which is smaller than the dynamical branching ratio of the two reaction pathways. 

Next, we examine the reaction dynamics inside the cavity by varying the cavity frequency $\omega_{\rm c}$ across the typical vibrational frequency range from $300$ to $2000~{\rm cm}^{-1}$. \Fig{fig3:model1}a) shows the ratios of reaction rates inside and outside the cavity ($k_{\rm f, \bf 1}/k^{\circ}_{\rm f, \bf 1}$ in red plotted against the left axis, and $k_{\rm f, \bf 2}/k^{\circ}_{\rm f, \bf 2}$ in green against the right axis) as a function of the cavity frequency, with a fixed light-matter coupling strength $\eta_{\mathrm{c}}=0.00125~$a.u.

For the first reaction pathway, the reaction rate inside the cavity experiences a pronounced resonant enhancement that peaks around $\omega_{\mathrm{c}}=780\mathrm{cm}^{-1}$. This arises from the resonant cavity-induced vibrational heating from the vibrational ground state to the first excited state, where the cavity frequency closely matches the corresponding energy gap $\Delta E_{0-1}=794~\mathrm{cm}^{-1}$ of the bare molecule. The peak broadening and redshifting are attributed to the broadening effects from coupling to the solvent, cavity mode, and cavity bath. Interestingly, the second reaction pathway is also enhanced in the low-frequency regime around $\omega_{\mathrm{c}}=780\mathrm{cm}^{-1}$ due to the broadening. However, in this case, we still observe a slight increase in the branching ratio $\phi$ of two reactive pathways, compared to the value $\phi^{\circ}$ outside the cavity. As the cavity frequency is further tuned up, while $k_{\rm f, \bf 1}$ decreases from its peak value, the rate constant for the second reaction pathway inside the cavity reaches its maximum at around $\omega_{\rm c}=1035~{\rm cm}^{-1}$, which agrees roughly with the transition energy $\Delta E_{0-2}=1051~\mathrm{cm}^{-1}$ from the vibrational ground state to the second excited state. At this frequency, the branching ratio $\phi=k_{\rm f, \bf 1}/k_{\rm f, \bf 2}$ reduces to 3.2.   

Moreover, both rate modification profiles for the two reaction pathways exhibit a secondary peak near $\omega_{\mathrm{c}}=1760~\mathrm{cm}^{-1}$. This subtle feature arises from the cavity-modified vibrational transition $|v_{\rm m}=0\rangle \rightarrow |v_{\rm m}=3\rangle$, which, while dominating the linear absorption spectrum, participates only in the slower, less favorable two-step channels of both reaction pathways. The disparity between spectroscopic prominence and dynamical rate modification intensity highlights a critical insight: chemical dynamics are influenced by factors that extend beyond those reflected in linear spectroscopy and can be governed by processes that are effectively spectroscopically silent.

The equilibrium populations of the two products as a function of the cavity frequency $\omega_{\mathrm{c}}$ for the reaction inside the cavity are displayed in \Fig{fig3:model1} b). Remarkably, in contrast to the rate alterations, which are most pronounced in the resonant conditions, the equilibrium populations of products ${\bf 1}$ and  ${\bf 2}$ remain largely unchanged, compared to the values outside the cavity. There is no notable characteristic feature in the equilibrium populations as a function of the cavity frequency.  Furthermore, this invariance in equilibrium distribution is robust and persists regardless of the specific form of the dipole function $\mu(x_{\rm m)}$.

\begin{figure}
\centering
  \begin{minipage}[c]{0.45\textwidth}
  \raggedright  a) 
    \includegraphics[width=\textwidth]{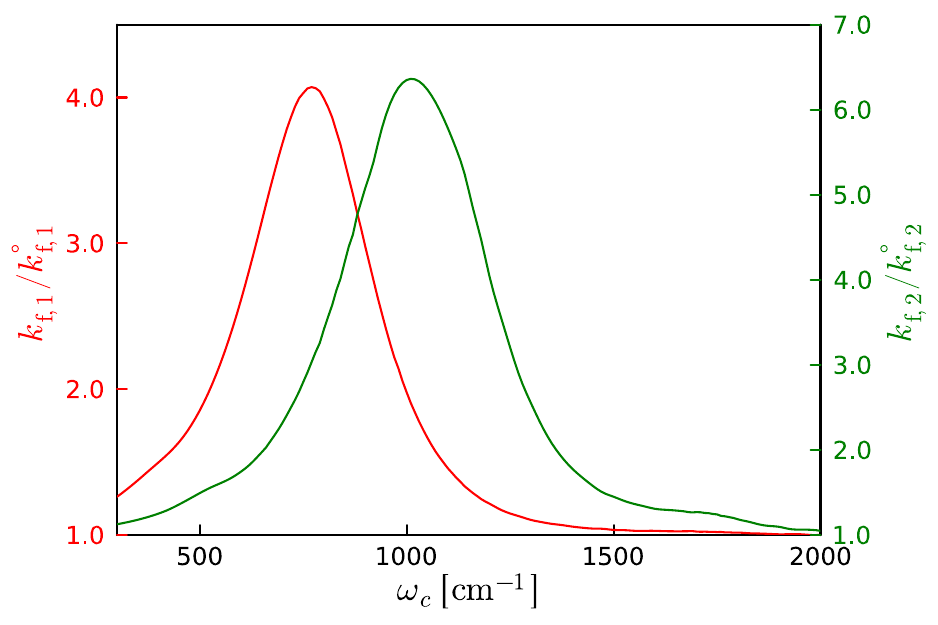}
  \end{minipage} 
\begin{minipage}[c]{0.45\textwidth}
  \raggedright  b) 
    \includegraphics[width=\textwidth]{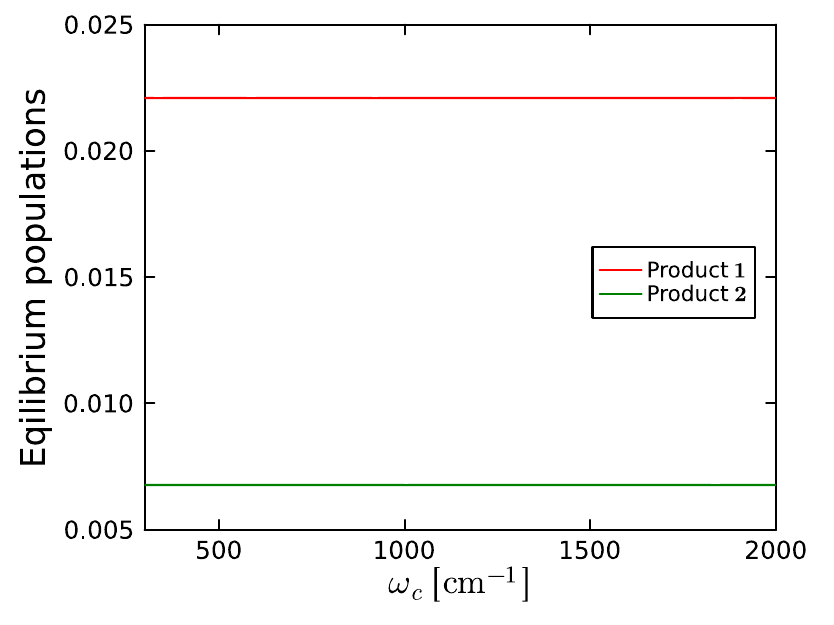}
  \end{minipage}
\caption{Same as \Fig{fig3:model1} but for a larger light-matter coupling strength $\eta_{\rm c}=0.005~$a.u.} \label{fig4:model1}
\end{figure}
The results for a larger light-matter coupling strength $\eta_{\mathrm{c}}=0.005~$a.u. are shown in \Fig{fig4:model1}. The reaction rate modification profiles are similar to those in \Fig{fig3:model1} a). However, both major resonant rate modification peaks corresponding to the two reaction pathways are significantly intensified and slightly red-shifted by roughly $10~{\rm cm}^{-1}$. In addition, the high-frequency side peaks at $\omega_{\rm c}=1760~{\rm cm}^{-1}$ are smoothed out. This suggests that cavity-induced resonant catalytic effects become stronger with an increased light-matter coupling strength. Notably, near $\omega_c=1020~{\rm cm}^{-1}$, the resonant frequency region of the second reaction pathway, the reaction rate $k_{\rm f,\bf 2}$ increases by a factor of 6.4, and the branching ratio $\phi=k_{\rm f, \bf 1}/k_{\rm f,\bf 2}$ decreases to 1.2. This exemplifies the cavity-induced selective excitation of vibrational modes. As before, the equilibrium populations inside the cavity are preserved, as shown in \Fig{fig4:model1} b). 

This finding is consistent with the conclusions of Refs.~\onlinecite{CamposGonzalezAngulo_2020_JCP_p,Li_2020_JCP_p234107}, which argue that classical transition state theory is inadequate for capturing the ground-state chemical reactivities observed in experiments. In this context, when the cavity frequency is tuned in resonance with specific molecular vibrational transition energies, the significant rate enhancement caused by the light-matter hybridization is attributed to dynamical non-equilibrium effects, which modify vibrational transition rates, rather than altering equilibrium properties. A similar observation is also made in an asymmetric double-well model, which is discussed in more detail in the Appendix A. 

\subsection{System II: ultrafast reaction dynamics}
\begin{figure}
\centering
  \begin{minipage}[c]{0.45\textwidth}
  \raggedright a) 
    \includegraphics[width=\textwidth]{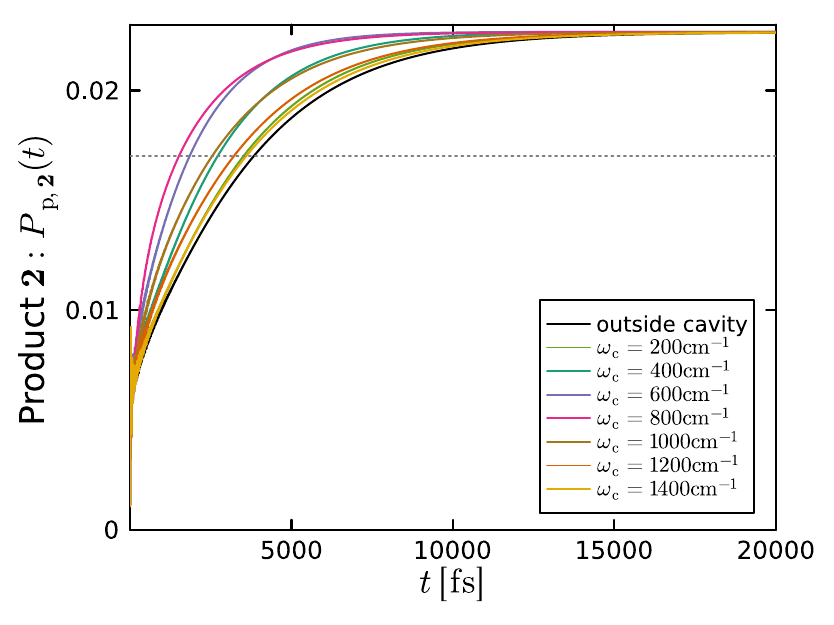}
  \end{minipage}
      \begin{minipage}[c]{0.45\textwidth}
  \raggedright b) 
    \includegraphics[width=\textwidth]{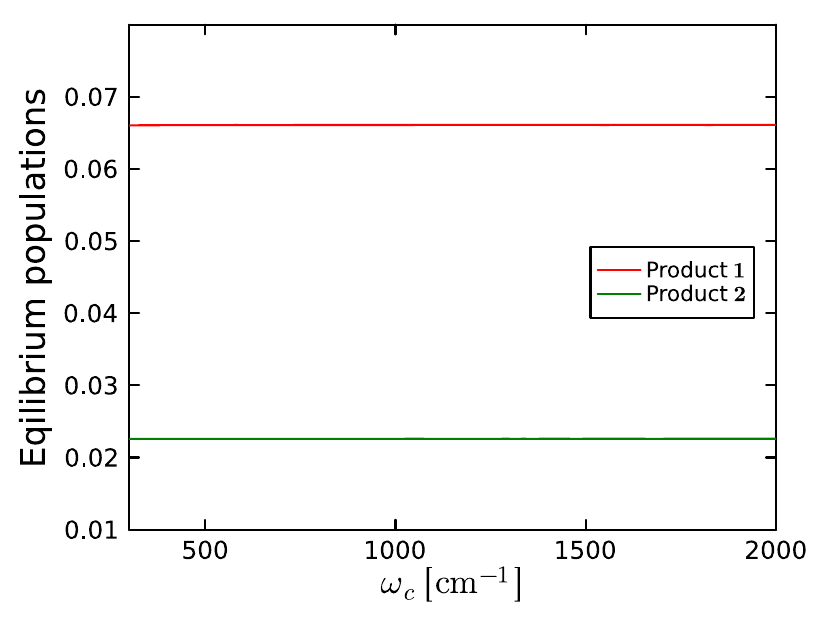}
  \end{minipage}
    \begin{minipage}[c]{0.45\textwidth}
  \raggedright c) 
    \includegraphics[width=\textwidth]{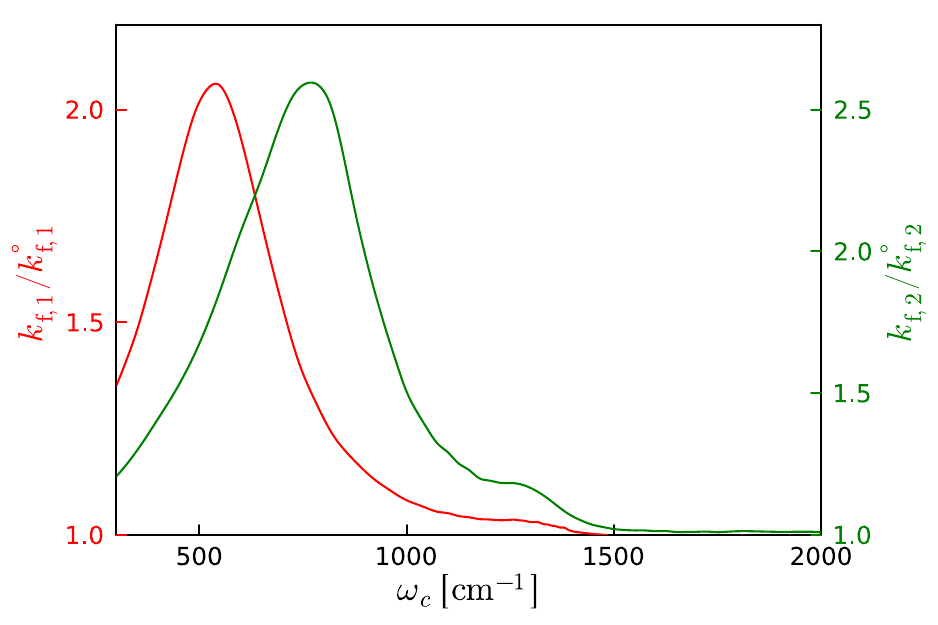}
  \end{minipage}
\caption{a) Population dynamics of product ${\bf 2}$, $P_{\rm p, \bf 2}(t)$, in System II for the reaction outside the cavity (black line) and inside the cavity at different frequencies (colored lines). b) Equilibrium population of two products, $P_{\rm p, \bf 1}^{\rm eq}$ and $P_{\rm p, \bf 2}^{\rm eq}$, as a function of $\omega_{\rm c}$. c) Ratios of forward reaction rates $k_{\rm f, \bf i}/k_{\rm f, \bf i}^{\circ}$ as a function of the cavity frequency $\omega_{\mathrm{c}}$, with red and green lines representing the first and second reaction pathways, respectively. The light-matter coupling strength is set to $\eta_{\mathrm{c}}=0.005~$a.u. 
} \label{fig5:model2}
\end{figure}
To gain a more intuitive understanding of the dynamical variations in reaction rates and the invariance of equilibrium populations, we consider an alternative parameterization of the PES with $E_0=15000~{\rm cm}^{-1}$, as shown in \Fig{fig1:PES}b).
In this scenario, the reaction barriers are significantly lower, leading to ultrafast reaction dynamics that complete within tens of picoseconds. 

As an illustrative example, \Fig{fig5:model2}a) displays the population dynamics of product ${\bf 2}$ outside the cavity (black line) and inside the cavity for different cavity frequencies (colored lines). We observed that, despite the variations in reaction speed, the long-time population inside the cavity converges to the same equilibrium as the field-free case, irrespective of cavity frequency. This long-time distribution is consistent with the equilibrium results obtained from imaginary-time propagation, as depicted in \Fig{fig5:model2}b). However, the reaction proceeds at different rates toward equilibrium, highlighting again the role of the cavity in modifying transient dynamics without altering equilibrium properties.

Unlike the previous model system where the separation of timescales $t_{\rm plateau} \gg t_{\rm transient}$ justifies the use of a well-defined rate expression in \Eq{rateexpression}, this assumption no longer holds in the present case. Thereby, the reaction dynamics do not strictly obey the kinetic equations given by \Eq{rateequation}. To quantify the reaction speed in this scenario, we introduce an ad hoc rate definition: $k_{\rm f,\bf 1}=1/t_{75\%,\bf 1}$ and $k_{\rm f,\bf 2}=1/t_{75\%,\bf 2}$, where $t_{75\%,\bf 1}$ and $t_{75\%, \bf 2}$ specify the time at which products ${\bf 1}$ and ${\bf 2}$ reach 75\% of their respective equilibrium populations, as illustrated by the gray dotted line in \Fig{fig5:model2}a). 

In the absence of the cavity, the reaction along the first and second pathways reaches 75\% completion at $t_{75\%,\bf 1}=3015~{\rm fs}$ and $t_{75\%,\bf 2}=3825~{\rm fs}$, respectively. This yields reaction rates of $k_{\rm f, \bf 1}=3.3\times10^{-4}~{\rm fs}^{-1}$ and $k_{\rm f, \bf 2}=2.6\times10^{-4}~{\rm fs}^{-1}$ with a branching ratio of $\phi=1.3$, indicating a slightly favored first reaction pathway. 

\Fig{fig5:model2}c) presents the ratios of reaction rates inside and outside the cavity for both reaction pathways as a function of the cavity frequency. The light-matter coupling strength is fixed at $\eta_{\mathrm{c}}=0.005~$a.u. While some trends resemble those observed in System I (see \Sec{subsec:modelI}), distinct differences also exist.  The rate modification profile for the first reaction pathway exhibits a peak centered at $\omega_{\rm c}=540~\mathrm{cm}^{-1}$, slightly redshifted relative to the vibrational transition energy $\Delta E_{0-1}$  from the ground state to the first excited state of the bare molecule. Meanwhile, tuning the cavity frequency is to the proximity of  $\omega_{\rm c}=770~\mathrm{cm}^{-1}$ enhances the second reaction pathway rate $k_{\rm f, \bf 2}$ by a factor of 2.6. This enhancement results from cavity-assisted vibrational heating, which facilitates dynamical population transfer from the vibrational ground state to the second excited state.  Under this condition, the branching ratio inside the cavity reverses to $\phi=0.67$, indicating that the second reaction pathway proceeds more rapidly. In addition, a shoulder feature at approximately $1300~\mathrm{cm}^{-1}$ corresponds to the indirect reaction pathways mediated by excitations to the third vibrationally excited state in the reactant region.

These results demonstrate that optical cavities can impact not only slow rate processes but also ultrafast reaction dynamics, giving rise to resonant rate modifications. More importantly, they reinforce our finding that the changes in chemical reactivity within polaritonic vibrational chemistry can be driven purely by dynamical non-equilibrium effects, without altering equilibrium properties. 

As a final remark, for ultrafast reactive processes, multidimensional spectroscopy may provide a more practical and informative means to probe transient dynamics, offering deeper and richer insight into quantum coherence effects.\cite{mukamel1995principles} Such approaches could significantly deepen our fundamental understanding of the dynamical origins underlying cavity-induced rate modifications. Extension toward this promising direction is left for future work.

\section{\label{sec:conclusion}Conclusion}
In summary, we present in this work a proof-of-concept study on selective resonant catalysis of condensed-phase chemical reactions within an optical microcavity, using a triple-well model that enables two competing reaction pathways.  We investigate both rate modifications and equilibrium properties in the single-molecule strong coupling limit through a fully quantum-mechanical treatment. To achieve this, we employ the numerical exact real- and imaginary-time HEOM method, combined with efficient tree tensor network algorithms. 

Our results reveal that light-matter coupling influences both ultrafast dynamical processes and slow rate kinetics. The most pronounced rate modifications occur when the cavity frequency resonates with a specific vibrational transition that immediately leads to product formation--even when this transition does not dominate the molecular absorption spectrum. It is also possible to alter reaction rates by tuning the cavity frequency to a vibrational transition within the reactant region, which is followed by a subsequent transition to produce the final product. Importantly, in the single-molecule limit with a non-vanishing light-matter coupling strength, we found that equilibrium populations remain largely unchanged, as compared to those outside the cavity, regardless of whether the cavity is tuned on- or off-resonance, contrasting sharply with the observed variations in reaction rates. These findings underscore the inherently non-equilibrium nature of cavity-induced modifications to chemical reactivities in resonant conditions, highlighting the necessity of dynamical simulations for accurate theoretical descriptions.

We emphasize that the present study is restricted to the single-molecule regime. In contrast, most current experimental realizations typically involve microcavities containing millions of molecules. However, it should not be assumed a priori that collective effects--wherein the cavity mediates interactions among molecules and alters their individual chemical reactivities beyond the single-molecule limit--automatically or universally emerge in such systems. In our prior work,\cite{Ke_J.Chem.Phys._2024_p224704} we examined a heterodimer composed of two molecules with non-overlapping vibrational transition energies, each strongly coupled to the cavity field. Due to the energetic mismatch, cavity-mediated intermolecular energy transfer was effectively blocked, and no collective enhancement of individual reaction rates was observed beyond those of the isolated monomers. Even in homogeneous systems where all molecules are identical, molecular aggregation via the cavity mode does not always play a key role in rate modifications.  For instance, in the presence of strong cavity damping, any energy transferred from one molecule to the cavity can rapidly dissipate to the cavity bath before it can be relayed to other molecules coupled to the same mode. In such scenarios, increasing the number of molecules has little impact on individual reaction rates, compared to the single-molecule case.\cite{Ke_J.Chem.Phys._2025_p64702} These findings suggest that the emergence and magnitude of collective effects depend sensitively on multiple factors, including the energetic alignment of molecular vibrational transitions, cavity loss rates, and light–matter coupling strength. A more systematic investigation across broader parameter regimes and diverse molecular models is essential to elucidate the precise conditions under which collective behavior becomes significant, how it scales with molecular number, and how it is modulated by environmental dissipation, particularly in the weakly damped cavity limit.

Moreover, further extension to the collective regime must also account for static collective factors, such as cavity-induced changes in molecular polarizabilities, altered molecular distributions, and modified local dipole-dipole interactions due to image charge effects.  These effects may significantly impact equilibrium properties. Recent analytical work has shown that electronic molecular polarizabilities can be modified even with vanishingly small single-molecule couplings.\cite{horak2025analytic} Future studies, along with system-specific ab initio simulations, are needed to fully capture these effects.

\begin{acknowledgments}
The author thanks the Swiss National Science Foundation for the award of a research fellowship (Grant No. TMPFP2\_224947). 
\end{acknowledgments}


\section*{Data Availability Statement}
The data and code that support the findings of this work are available from the corresponding author upon reasonable request.

\appendix
\section{\label{appendixa}Asymmetric double-well model}
\begin{figure}
\centering
  \begin{minipage}[c]{0.45\textwidth}
  \raggedright a) 
    \includegraphics[width=\textwidth]{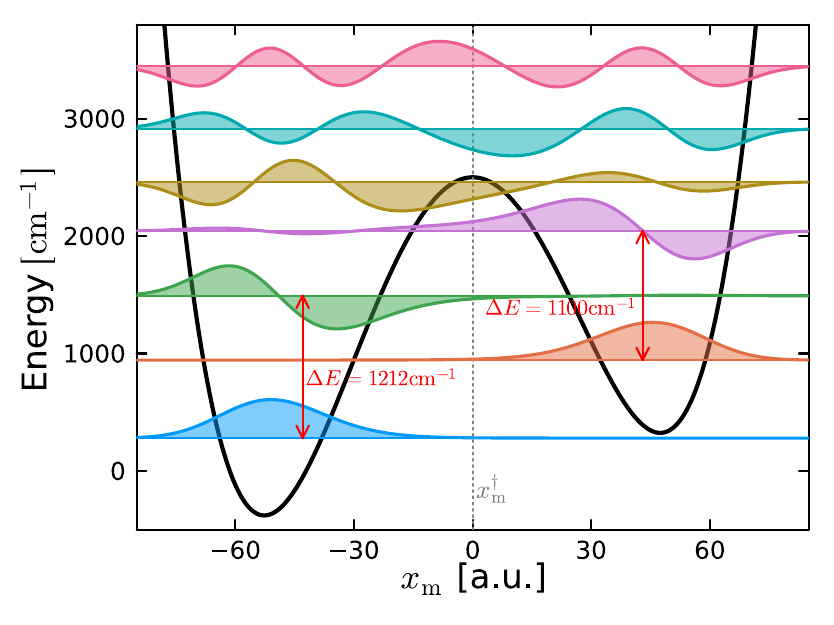}
  \end{minipage}
    \begin{minipage}[c]{0.45\textwidth}
  \raggedright b) 
    \includegraphics[width=\textwidth]{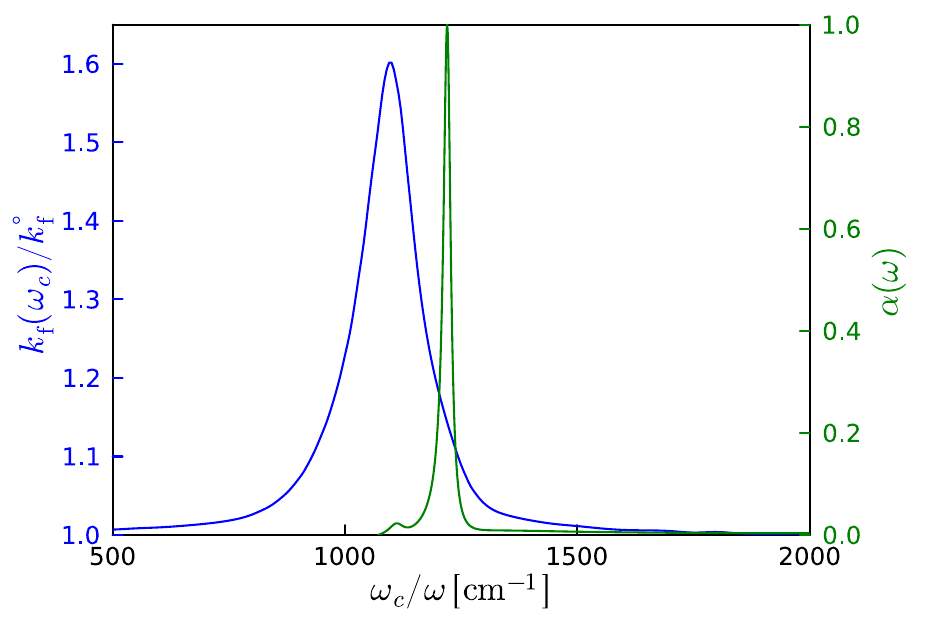}
  \end{minipage}
    \begin{minipage}[c]{0.45\textwidth}
  \raggedright c) 
    \includegraphics[width=\textwidth]{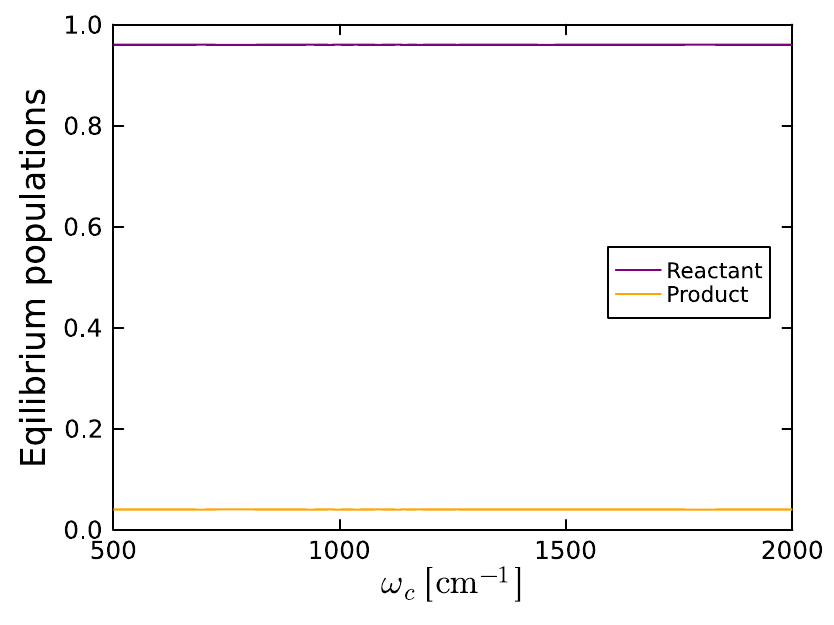}
  \end{minipage}
\caption{a) Potential energy surface for an asymmetric double-well model, as defined in \Eq{PES_asymmetric}, with parameters $E_{0}=2500~\mathrm{cm}^{-1}$, $a=50~\mathrm{a.u.}$, and $c=350~\mathrm{cm}^{-1}$. The dividing surface is placed at $x_{\rm m}^{\ddagger}=0$, indicated by the dotted gray line. b) Ratios of forward reaction rates from the left well to the right well inside and outside the cavity, $k_{\rm f}/k_{\rm f}^{\circ}$, as a function of the cavity frequency $\omega_{\mathrm{c}}$ (blue line, left axis), where $k_{\rm f}^{\circ}$ is the field-free rate. For comparison, the molecular absorption profile outside the cavity is shown (green line, right axis). Equilibrium population of the reactant and product in the asymmetric double-well system. The light-matter coupling strength is fixed at $\eta_{\mathrm{c}}=0.00125~$a.u.} \label{figS:model3}
\end{figure}

In this appendix, we explore the reaction dynamics and thermal equilibrium of an asymmetric double-well model inside an optical cavity in the single-molecule limit. The potential energy surface is given by\cite{Lindoy_2023_NC_p2733}
\begin{equation}
\label{PES_asymmetric}
    U_{\rm DW}(x_{\rm m}) = E_{\mathrm{0}}\left(\left(\frac{x_{\rm m}}{a}\right)^2-1\right)^2+c\left(\frac{x_{\rm m}}{a}\right)^3.
\end{equation}
where we set $E_{0}=2500~\mathrm{cm}^{-1}$, $a=50~\mathrm{a.u.}$, and $c=350~\mathrm{cm}^{-1}$, and it is depicted in \Fig{figS:model3}a). The dividing surface is placed at $x_{\rm m}^{\ddagger}=0$, which separates the reactant and product regions. The lowest eigenstate is exclusively localized in the left well, $x_{\rm m}<x_{\rm m}^{\ddagger}$, which is defined as the reactant region. Starting from the ground state, excitation to the second vibrationally excited state $|v_{\rm m}=2\rangle$ is most probable due to its dominant transition dipole moment, i.e., $|\langle v_{\rm m}=2|x_{\rm m}|v_{\rm m}=0\rangle|\gg |\langle v_{\rm m}\neq 2|x_{\rm m}|v_{\rm m}=0\rangle|$, with transition energy of $\Delta E_{0-2}=1212~{\rm cm}^{-1}$. The first and third vibrationally excited states ($|v_{\rm m}=1\rangle$ and $|v_{\rm m}=3\rangle$) are primarily localized in the product region, as shown in \Fig{figS:model3}a), and their energy gap is $\Delta E_{1-3}=1100~{\rm cm}^{-1}$.  The fourth excited state is near the reaction barrier and has a non-negligible population probability in both wells. 

For this system, the reaction dynamics are governed by rate processes. The rigorous expression for the first-order forward reaction rate constant, which describes the transition from the reactant (left well) to the product (right well), is given in the reactive flux formalism as\cite{Miller_1983_JCP_p4889,Craig_2007_JCP_p144503,Chen_2009_JCP_p134505,Ke_2022_JCP_p34103}
\begin{equation}
k_{\rm f}=\lim_{t\rightarrow t_{\rm plateau}} \frac{\mathrm{tr}\{\rho(t)F\}}{1-(1+P_{\rm r}^{\rm eq}/P_{\rm p}^{\rm eq})P_{\rm p}(t)}.
\end{equation}
Here, the flux operator is defined as $F=i[H, h]$, where $h=\theta(x-x_{\rm m}^{\ddagger})$ is the side operator projecting onto the product region. The time-dependent population in the product region is given by $P_{\rm p}(t)=\mathrm{tr}\{h\rho(t)\}$, while  
$P_{\rm r}^{\rm eq}=\mathrm{tr}\{(1-h)\rho^{\rm eq}(\beta)\}$ and $P_{\rm p}^{\rm eq}=\mathrm{tr}\{h\rho^{\rm eq}(\beta)\}$ represent the equilibrium populations of the reactant and product, respectively. The time $t_{\rm plateau}$ denotes when the reaction rate reaches a plateau value. The simulations are performed with the following parameters: $\lambda_{\rm m}=\Omega_{\rm m}=100~{\rm cm}^{-1}$, $\lambda_{\rm c}=150~{\rm cm}^{-1}$, $\Omega_{\rm c}=1000~{\rm cm}^{-1}$. 

By carrying out the vibrationally resolved calculations of the field-free forward reaction rate outside the cavity $k_{\rm f}(d_{\rm m})$, we identified two reaction channels. The primary channel, $|v_{\rm m}=0\rangle\rightarrow |v_{\rm m}=2\rangle\rightarrow|v_{\rm m}=3\rangle\rightarrow|v_{\rm m}=1\rangle$ yields a rate of $k_{\rm f}(d_{\rm m}=4)=4.5\times 10^{-8}~{\rm fs}^{-1}$.  Notably, both intra-well transitions, $|v_{\rm m}=0\rangle\rightarrow |v_{\rm m}=2\rangle$ and  $|v_{\rm m}=1\rangle\rightarrow |v_{\rm m}=3\rangle$, appear as distinct peaks in the molecular absorption spectrum, as illustrated in \Fig{figS:model3}b). However, the former transition, which originates from the vibrational ground state on the reactant side, exhibits significantly stronger absorption intensity due to the higher thermal population of $|v_{\rm m}=0\rangle$. In contrast, the product side transition, although having a slightly larger dipole matrix element ($|\langle v_{\rm m}=3|x_{\rm m} |v_{\rm m}=1\rangle|=9.47$ vs. $|⟨v_{\rm m}=2|x_{\rm m}| v_{\rm m}=0\rangle|=9.25$), appears only as a weak secondary absorption peak owing to the lower thermal population of the first vibrational excited state. 
Additionally, higher-lying excited states, particularly $|v_{\rm m} =4\rangle$, which are spectroscopically dark, also contribute despite playing a minor role in the overall reaction rate. The fully converged forward reaction rate outside the cavity is $k_{\rm f}^{\circ}=5.0\times 10^{-8}~{\rm fs}^{-1}$. 

To examine the impact of the cavity mode, we set the light-matter coupling strength to $\eta_{\rm c}=0.00125~{\rm a.u}$ and calculate the modified reaction rates inside the cavity $k_{\rm f}$ with varying cavity frequency $\omega_{\rm c}$. The ratios, $k_{\rm f}/k_{\rm f}^{\circ}$ for the forward reaction, are plotted in \Fig{figS:model3}b) as a function of the cavity frequency. The rate modification profile exhibits a sharp peak centered around $\omega_{\rm c}=1090~{\rm cm}^{-1}$, which is resonant with the vibrational transition in the product region between the first and third vibrationally excited states. Interestingly, despite the strong absorption strength of the reactant-side vibrational transition $|v_m=0\rangle \rightarrow |v_m=2\rangle$, no corresponding peak appears in the rate modification profile. This can be rationalized by recognizing that the $|v_m=0\rangle \rightarrow |v_m=2\rangle$ transition is a fast, non-rate-limiting step, whereas the $|v_m=3\rangle \rightarrow |v_m=1\rangle$ transition constitutes the slower, rate-limiting step in the reaction pathway.  When the cavity is tuned to $\omega_{\rm c}=1100\,\mathrm{cm}^{-1}$ resonance, the light field accelerates this bottleneck cooling step $|v_m=3\rangle \rightarrow |v_m=1\rangle$, thereby substantially enhancing the overall reaction rate. Comparatively, resonantly enhancing the faster, upstream transition $|v_m=0\rangle \rightarrow |v_m=2\rangle$ in the reaction region leads to only marginal changes in the overall rate. This finding challenges the assumption underlying the previous analytical studies \cite{Ying_2023_JCP_p84104,Ying_2024_CM_p110} that the reactant-side transition is always the rate-limiting step.  
Nevertheless, a deeper understanding of why the product-side transition emerges as rate-limiting still merits further investigation. Factors such inter-state couplings,  the structure of the transition dipole operator, and solvent-induced frictions may subtly shift the kinetic bottleneck in this and similar systems. A broader, systematic survey of reaction potentials, dipole functions, and environmental parameters will be essential for identifying and predicting the rate-controlling step across diverse molecular systems--an endeavor that will ultimately facilitates more accurate and generalizable studies in polariton chemistry.

More importantly, this example further underscores the greater complexity of reaction dynamics, which cannot be fully deduced from spectroscopic signatures alone. It highlights that dynamical factors beyond transition dipole strengths and thermal populations--such as the relative timescale of individual reaction steps--also play a crucial role in determining whether, under what conditions, reaction rate modifications can be observed inside an optical cavity.

In contrast, the equilibrium populations of the reactant and product (or equivalently the chemical equilibrium constant $\chi=P_{\rm r}^{\rm eq}/P_{\rm p}^{\rm eq}$) remain constant across different cavity frequencies, as shown in \Fig{figS:model3}c). This result aligns with our findings from the triple-well model, confirming that cavity-induced resonant modifications to chemical reactivity originate from dynamical and non-equilibrium effects rather than changes in equilibrium properties.
Consequently, interpreting experimental data through classical transition state theory, which attributes rate modifications to the alterations in the molecular energy landscape, should be approached with caution.

%

\end{document}